\documentclass[conference]{IEEEtran}
\IEEEoverridecommandlockouts
\usepackage{cite}
\usepackage{amsmath,amssymb,amsfonts}
\usepackage{algorithmic}
\usepackage{graphicx, booktabs}
\usepackage{xcolor}
\usepackage{tikz}
\usepackage{hyperref}
\hypersetup{
    colorlinks=true,
    linkcolor=blue,
    filecolor=black,      
    urlcolor=black,
    pdftitle={IEEE-IS² 2024 Music Packet Loss Concealment Challenge},
    }
\usepackage{xurl}
\usepackage[caption=false]{subfig}
\usepackage{flushend}
\def\BibTeX{{\rm B\kern-.05em{\sc i\kern-.025em b}\kern-.08em
    T\kern-.1667em\lower.7ex\hbox{E}\kern-.125emX}}

\begin{document}
\title{The IEEE-IS² 2024 Music Packet Loss Concealment Challenge}

\author{\IEEEauthorblockN{Alessandro Ilic Mezza}
\IEEEauthorblockA{\textit{Dipartimento di Elettronica, Informazione e Bioingegneria} \\
\textit{Politecnico di Milano}\\
Milan, Italy \\
\texttt{alessandroilic.mezza@polimi.it}}
\and
\IEEEauthorblockN{Alberto Bernardini}
\IEEEauthorblockA{\textit{Dipartimento di Elettronica, Informazione e Bioingegneria} \\
\textit{Politecnico di Milano}\\
Milan, Italy \\
\texttt{alberto.bernardini@polimi.it}}
}

\maketitle

\begin{abstract}
We present the IEEE-IS² 2024 Music Packet Loss Concealment Challenge. We begin by detailing the challenge rules, followed by an overview of the provided baseline system, the blind test set, and the evaluation methodology used to determine the final ranking. This inaugural edition aimed to foster collaboration between researchers and practitioners from the fields of signal processing, machine learning, and networked music performance, while also laying the groundwork for future advancements in packet loss concealment for music signals.
\end{abstract}


\section{Introduction}
Packet loss, either by missing packets or high packet jitter, is one of the main problems and, in turn, engineering challenges for real-life Networked Music Performance (NMP) applications. While Packet Loss Concealment (PLC) for Voice over IP has recently attracted a great deal of attention, as also evidenced by the recent INTERSPEECH 2022 Audio Deep Packet Loss Concealment Challenge \cite{diener2022interspeech} and the 2024 ICASSP Audio Deep Packet Loss Concealment Grand
Challenge \cite{diener2024icassp}, PLC for NMP has been considerably less studied.

The IEEE-IS² 2024 Music Packet Loss Concealment Challenge\footnote{Challenge web page: \href{https://internetofsounds2024.ieee-is2.org/program/ieee-is\%C2\%B2-2024-music-packet-loss-concealment-challenge}{https://internetofsounds2024.ieee-is2.org/program/\\ieee-is²-2024-music-packet-loss-concealment-challenge}} aimed to provide a platform for researchers and practitioners working on the topic to share their work and compare different methods within a unified benchmark, in an effort to encourage further advancements in the field of Music PLC.

\section{Challenge Overview}
\subsection{Challenge Rules}
The IEEE-IS² 2024 Music Packet Loss Concealment Challenge lunched on May 13, 2024; the blind test set was released on July 3, 2024; the submission window closed on July 20, 2024. Each team was allowed to submit up to two systems for evaluation.

The systems had to be designed to process audio files at a sampling rate of $44.1$~kHz, and predict packets of $512$ samples, corresponding to approximately $11.6$~ms.

Whereas smaller packets are sometimes preferred in NMP application, %as they depend on the buffer size of the soundcard from which audio data is read, 
the choice of using packets of $512$ samples was meant to be challenging for the proposed PLC systems and encourage the participants to tackle harder test cases with long-term losses.

Additionally, motivated by the tight latency requirements of real-time networked musical interaction, only causal systems were deemed eligible for the Challenge. Namely, at any given time, only previously received packets or prediction thereof may be used to predict the next audio frame. In other words, differently from other audio deep PLC challenges, systems were not allowed any look ahead.
Other than that, there were no limitations on the eligible PLC methods, which may comprise one or more deep-learning models, traditional signal processing algorithms, or a hybrid approach.

We did not provide training data, nor did we indicate a list of eligible training datasets.
However, the Challenge prescribed that participants only used data from publicly-available and freely-accessible datasets.
No limit, instead, was posed on data augmentation, as long as the models were kept blind to metadata and other auxiliary information other than packet loss traces.

We encouraged all participants to develop PLC systems that would respect real-time constraints as strictly as possible. However, slower-than-real-time inference was not accounted as a reason for disqualifying a submission.

\begin{table}[t]
    \centering
    \caption{Challenge Ranking}
    \label{tab:ranking}
    \resizebox{\linewidth}{!}{%}
    \begin{tabular}{lcccc}
    \toprule
        & Average score{\tiny\ ± SD} & Median score & Trials won & Ranking\\
        \midrule
        PARCnet-IS² (Baseline) & $58.06$\,${\scriptscriptstyle \pm\, 22.13}$ & $59.5$ & $9$ & $1$\textsuperscript{st}\\
        Aironi et al.\ (full) \cite{aironi:challenge} & $49.14$\,${\scriptscriptstyle \pm\, 22.09}$ & $50.5$ & $1$ & $2$\textsuperscript{nd}\\
        Aironi et al.\ (lite) \cite{aironi:challenge}  & $48.06$\,${\scriptscriptstyle \pm\, 22.48}$ & $50.0$ & -- & $3$\textsuperscript{rd}\\
        Daniotti et al.\ \cite{daniotti:challenge} & $41.70$\,${\scriptscriptstyle \pm\, 21.38}$ & $41.0$ & -- & $4$\textsuperscript{th}\\ 
        Zero-filling & $5.14$\,${\scriptscriptstyle \pm\, 8.55}$ & 0.0& -- & --\\ 
    \bottomrule
    \end{tabular}
    }
\end{table}

\subsection{Baseline System}

We released a baseline system for the IEEE-IS² 2024 Music Packet Loss Concealment Challenge. The system, dubbed \textbf{PARCnet-IS²}, is a modified PARCnet architecture \cite{mezza2024parcnet} trained on Medley-solos-DB \cite{medley-solos-db}.

PARCnet comprises two parallel modules, an autoregressive linear predictor (AR model) and a feed-forward neural network.
The linear predictor is fitted in real-time within a sliding context window using the autocorrelation method with white noise compensation, while the neural network is trained to estimate the residual of the AR model.
Compared to the original method, PARCnet-IS² incorporates several minor modifications. Namely, (i)~PARCnet-IS² was trained for 250,000 steps using a $L^1$-loss instead of a $L^2$-loss; (ii)~the audio signals were sampled at $44.1$ kHz instead of $32$ kHz; (iii)~the system is designed to predict packets of $512$ samples instead of $320$; (iv)~the neural network valid context was increased from $7$ to $8$ packets; (v)~the order of the parallel AR model was increased to $256$; (vi)~the extra prediction length, which allows to cross-fade between subsequent packets, either valid or predicted, was increased from $80$ to $256$ samples.
For more details, we refer the readers to~\cite{mezza2024parcnet}. 
PARCnet-IS² is available online.\footnote{Available: \url{https://github.com/polimi-ispl/2024-music-plc-challenge/tree/main/parcnet-is2}}

\subsection{Blind Test Set}

The IEEE-IS² 2024 Music Packet Loss
Concealment Challenge blind test set\footnote{Available: \url{https://github.com/polimi-ispl/2024-music-plc-challenge}} consists of $162$ single-channel audio files in a $16$bit-$44.1$kHz wav format extracted from AVAD-VR~\cite{thery2019anechoic}, a publicly available dataset of anechoic audio and 3D-video recordings of several small music ensemble performances. 
Every test audio file consists of a $11.6$-second clip of a closed-miked classical or jazz performance obtained by segmenting the full recording with no overlap. 
The blind set thus comprises various acoustic instruments, including violin, cello, clarinet, sax, double bass, and classical guitar. Clips in which silence made up more than $30\%$ of the total duration were discarded.

The audio clips are artificially degraded by dropping packets (zero-filling) according to predetermined ``packet traces,'' i.e., text files containing a string of binary digits: 0 if a packet was correctly received and 1 if the packet was lost. Every digit in a packet trace corresponds to $512$ samples. Traces do not contain explicit temporal information, and the packet rate is implicitly determined by the audio sampling rate.
The packet traces used to create the blind test set were repurposed from the blind set of the INTERSPEECH 2022 Audio Deep Packet Loss Concealment Challenge.\footnote{Available: \url{https://github.com/microsoft/PLC-Challenge}}

Said traces are measured and represent a real network scenario. The text files are divided into three subsets according to the maximum burst loss length:
\textbf{Subset 1.}\ bursts of up to $6$ consecutive packets;
\textbf{Subset 2.}\ bursts of $6$ to $16$ consecutive packets;
\textbf{Subset 3.}\ bursts of $16$ to $50$ consecutive packets.
We sampled packet traces from Subset 1 (with a probability of $90\%$) and Subset 2 (with a probability of $10\%$). We did not sample traces from Subset 3. For each audio clip, we sampled and concatenated up to three traces so as to encompass the entire clip duration.

The clean, untampered versions of the degraded audio clips in the test set were kept private, making it a \textit{blind} set; these files were used for both objective and subjective evaluations.

\subsection{Evaluation Procedure}
Challenge participants were asked to download the blind test set, process each and every clip with the proposed PLC method, and submit the enhanced audio files. Similarly to~\cite{diener2024icassp}, no model was collected and run during the evaluation process.

The Challenge ranking was determined through a Multiple Stimuli with Hidden Reference and Anchor (MUSHRA) listening test. 
A subset of ten audio files was manually selected from the blind test set so as to encompass different musical instruments and playing styles.

The MUSHRA test thus consisted of ten \textit{trials}, where the test conditions were compared with the clean, untampered track in terms of Basic Audio Quality (BAQ). In each trial, the test condition that received the highest average score was considered the \textit{winner} of that trial. The final ranking was determined by counting the \textit{number of trials won} by each PLC method.
The average MUSHRA scores computed across all trials were considered as a tie-breaker.

\section{Team Submissions}
We received three systems from two teams. Additional information on each method can be found on the respective technical reports.\footnote{The technical reports are available at \href{https://internetofsounds.net/ieee-is\%C2\%B2-2024-music-packet-loss-concealment-challenge}{https://internetofsounds.net/ieee-is²-2024-music-packet-loss-concealment-challenge}}

\textbf{Daniotti et al.}~\cite{daniotti:challenge} submitted a variant of the original PARC-net model trained using a novel \textit{Tilt Loss}. This perceptually-motivated $L^1$-loss function adaptively reweights the frequency axis of the mel-spectrogram error emphasizing the high-frequency range, akin to an upward tilt filter.
%The proposed Tilt Loss is defined as $L_{1,1}$-norm of the adaptively reweighed difference between the predicted and ground truth 128-bin mel-spectrograms. The  spectral reweighting, acting as an upward tilt filter, applies a 1D soft mask to the mel-frequency bins of each short-time frame in order to adaptively emphasize high-frequency bands. 
The proposed PLC system was trained on the \textit{Bach Cello Suite} dataset \cite{chafe2020cello}.

\textbf{Aironi et al.}~\cite{aironi:challenge} proposed a novel PLC method that uses a bin2bin Generative Adversarial Network (GAN) \cite{aironi2023time} to generate audio conditioned by the estimate of a linear predictor. The bin2bin model is trained with a linear combination of spectral convergence, log-magnitude STFT loss, and least-square conditional GAN objectives.  
Aironi and colleagues submitted two systems, a \textit{full} model (54.4~M parameters) and a \textit{lite} model (3.4~M parameters), each trained on an ensemble of three datasets: Medley-solos-DB~\cite{medley-solos-db}, the Good-sounds.org dataset~\cite{good-sounds}, and 45~hours of MIDI clips from MAESTRO~\cite{hawthorne2018enabling} synthesized using SoundFonts. 

\section{Evaluation}

\subsection{Objective Evaluation}
\label{ssec:objective_eval}
Here, we provide a brief overview of the objective metrics considered as part of the systems evaluation. These metrics have been calculated on all clips in the blind test set. 
Even if prior studies have observed a statistically significant correlation between some objective metrics and subjective judgments~\cite{mezza2024parcnet}, no metric has been definitely proved to work for Music PLC algorithms. For this reason, the final ranking was only determined from the outcome of the MUSHRA test (Section \ref{ssec:subjective_eval}).

Let $y[n]$ and $\hat{y}[n]$ be the $N$-sample reference and enhanced waveforms, respectively. We also define the vectors $\mathbf{y}=[y[0], ..., y[N]]^T$ and $\hat{\mathbf{y}}=[\hat{y}[0], ..., \hat{y}[N]]^T$. We denote the $L^2$-norm with $\lVert \cdot \rVert_2$.

As far as time-domain metrics are concerned, we compute the Mean Squared Error~(MSE)
\begin{equation}
    \text{MSE} = \frac{1}{N}\sum_{n=0}^{N-1} \left(y[n] - \hat{y}[n]\right)^2\,,
\end{equation}
the Signal-to-Distortio Ratio~(SDR)
\begin{equation}
    \text{SDR} = 10\log_{10}\frac{\lVert \mathbf{y} \rVert_2^2}{\lVert \mathbf{y} - \hat{\mathbf{y}} \rVert_2^2}\,,
\end{equation}
and the Scale-Invariant SDR (SI-SDR)~\cite{le2019sdr}
\begin{equation}
    \text{SI-SDR} = 10\log_{10}\frac{\lVert \alpha \mathbf{y} \rVert_2^2}{\lVert \alpha\mathbf{y} - \hat{\mathbf{y}} \rVert_2^2}\,,
\end{equation}
where $\alpha=\hat{\mathbf{y}}^T\mathbf{y} / \lVert \mathbf{y} \rVert_2^2$.

\begin{table}[t]
    \centering
    \caption{Impairment description for PLCMOS and PEAQ scores.}
    \label{tab:impairment_description}
    \begin{tabular}{lcc}
    \toprule
    Impairment description	& PLCMOS & PEAQ ODG\\
    \midrule
    Imperceptible & $5.0$ & $\phantom{-}0.0$ \\
    Perceptible, but not annoying &	$4.0$ & $-1.0$\\
    Slightly annoying & $3.0$ & $-2.0$ \\
    Annoying & $2.0$ & $-3.0$\\
    Very annoying & $1.0$ & $-4.0$\\
    \bottomrule
    \end{tabular}
\end{table}

Additionally, we take into account metrics in the frequency and cepstral domain, respectively. Namely, we compute the Log-Spectral Distance (LSD)~\cite{gray1976distance}
\begin{equation}
    \text{LSD} = \frac{1}{M}\sum_{m=0}^{M-1}\sqrt{\frac{1}{K}\sum_{k=0}^{K} \log\lvert Y_m[k] \rvert^2 - \log\lvert \hat{Y}_m[k] \rvert^2}\,,
\end{equation}
where $\lvert Y_m[k] \rvert$ is the $(K+1)$-bin magnitude spectrum of the $m$-th reference signal frame $y_m[n]$ of length $2048$ samples, extracted using a Hann window with a hop size of $512$. 
Next, we compute the Mel-Cepstral Distance (MCD)~\cite{kubichek1993mdc}
\begin{equation}
    \text{MCD}= \frac{1}{M}\sum_{m=0}^{M-1}\sqrt{\sum_{i=1}^{16} \left(C_m[i] - \hat{C}_m[i]\right)^2}\,,
\end{equation}
where $C_m[i]$ is the $i$th MFCC extracted from the $1024$-sample frame $y_m[n]$ integrated over $20$ critical bands using a triangular mel-filterbank. Note that the zeroth coefficient was excluded~\cite{kubichek1993mdc}.

Furthermore, we compute an Objective Difference Grade (ODG) for each clip in the blind set with the ITU-R BS.1387 Perceptual Evaluation of Audio Quality (PEAQ) \cite{thiede2000peaq}. In particular, we use the open-source MATLAB implementation of the PEAQ Basic algorithm by P.~Kabal~\cite{kabal2002peaq}. For this metric, every clip had to be upsampled from $44.1$ to $48$~kHz.

Finally, we evaluate the latest version\footnote{Available: \url{https://pypi.org/project/speechmos}} of PLCMOS \cite{diener2023plcmos}, recently released for the ICASSP 2024 Audio Deep Packet Loss Concealment Grand
Challenge. Note that this data-driven non-intrusive metric was proposed for and trained with corrupted speech signals. Hence, it is unclear whether it is reliable when it comes to Music PLC. Moreover, PLCMOS is intended for signals at a sampling rate of $16$ kHz. Therefore, every clip had to be downsampled accordingly. Here, we report the PLCMOS results for completeness. 
The impairment descriptions for the PEAQ and PLCMOS scores are reported in Table~\ref{tab:impairment_description}.

\begin{figure}[t]
    \centering
    \includegraphics[width=\linewidth]{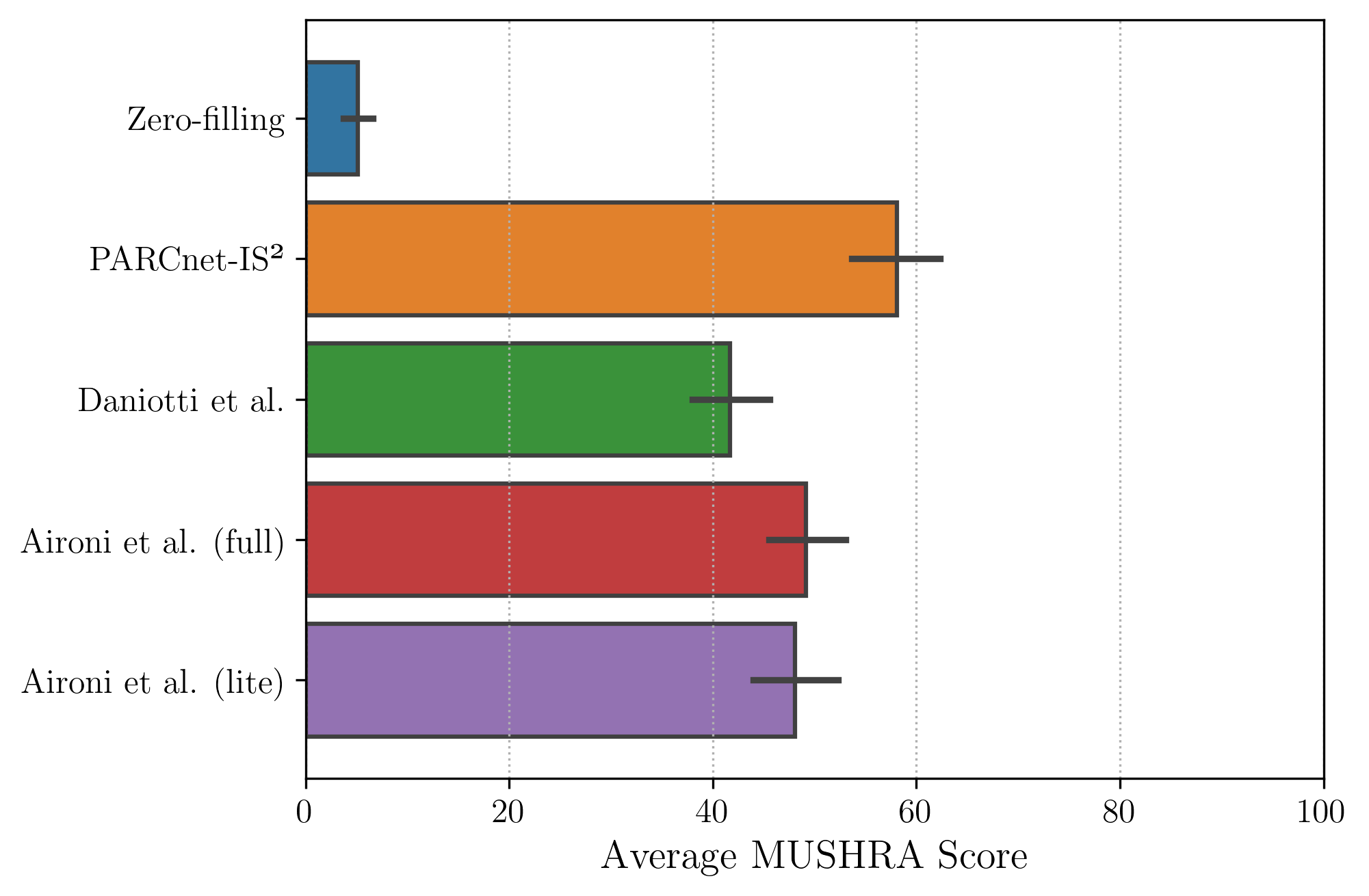}\label{fig:global_average_score}\\
    \vspace{1em}
    \includegraphics[width=\linewidth]{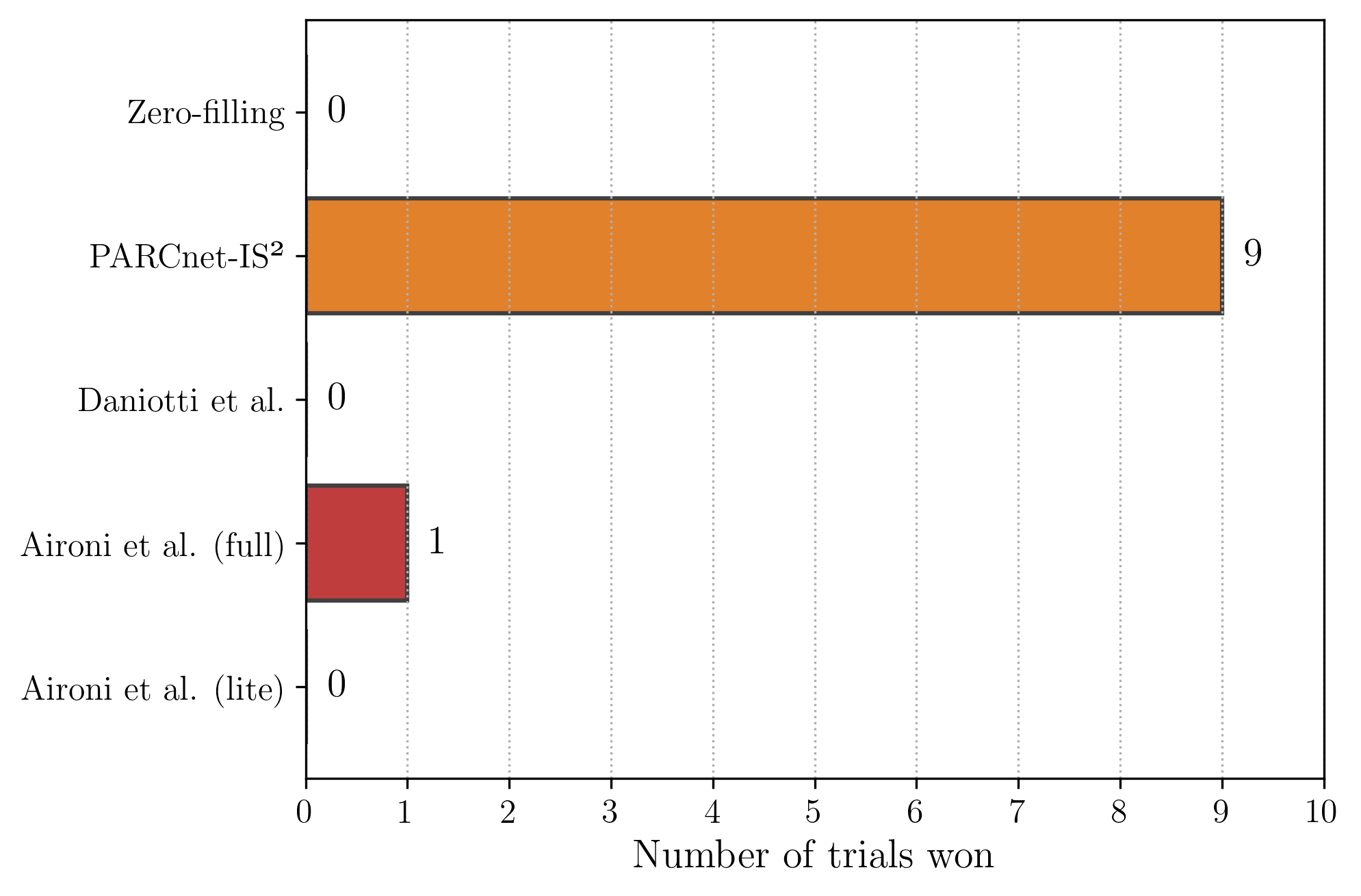}\label{fig:num_wins}\\
    \caption{Results of the MUSHRA test.}
    \label{fig:ranking}
\end{figure}

\begin{table*}[t]
    \centering
    \caption{Objective metrics computed on the entire blind set. Mean $\pm$ standard deviation; bold indicates the best value for each metric.\\$\uparrow$: higher is better; $\downarrow$: lower is better.}
    \label{tab:objective_metrics}
    \resizebox{\linewidth}{!}{%}
    \begin{tabular}{lccccccc}
    \toprule
    & \multicolumn{3}{c}{Time-domain} & Spectral & Cepstral & \multicolumn{2}{c}{Perceptual}\\
    \cmidrule(r){2-4} \cmidrule(lr){5-5} \cmidrule(lr){6-6} \cmidrule(l){7-8}
       & MSE$\times10^{-4}$ ($\downarrow$)   & SDR ($\uparrow$) & SI-SDR ($\uparrow$) & LSD ($\downarrow$) & MCD ($\downarrow$) & PEAQ ($\uparrow$) & PLCMOS ($\uparrow$) \\
    \midrule
    Zero-filling & $1.837 \pm 3.370$ & $11.87 \pm 4.07$ & $11.43 \pm 4.41$ & $0.599 \pm 0.477$ & $5.503 \pm 4.710$ & $-3.000 \pm 0.866$ & $1.717 \pm 0.414$ \\
    PARCnet-IS² (Baseline) & $\mathbf{0.645 \pm 1.096}$ & $\mathbf{16.33 \pm 6.49}$ & $\mathbf{17.95 \pm 6.45}$ & $\mathbf{0.239 \pm 0.158}$ & $\mathbf{2.456 \pm 2.062}$ & $\mathbf{-1.832 \pm 1.052}$ & $\mathbf{1.953 \pm 0.589}$ \\
    Daniotti et al.\ & $8.774 \pm 7.004$ & $13.21 \pm 8.51$ & $13.34 \pm 8.45$ & $0.279 \pm 0.211$ & $3.883 \pm 4.644$ & $-2.198 \pm 1.053$ & $1.935 \pm 0.496$ \\
    Aironi et al.\ (full) & $1.423 \pm 2.409$ & $13.32 \pm 5.14$ & $13.02 \pm 5.40$ & $0.290 \pm 0.189$ & $2.991 \pm 2.388$ & $-2.048 \pm 1.031$ & $1.849 \pm 0.510$ \\
    Aironi et al.\ (lite) & $1.307 \pm 2.189$ & $13.04 \pm 5.33$ & $13.03 \pm 5.35$ & $0.294 \pm 0.192$ & $3.100 \pm 2.476$ & $-2.069 \pm 1.031$ & $1.857 \pm 0.513$ \\
    \bottomrule
    \end{tabular}
}
\end{table*}

\subsection{Subjective Evaluation}
\label{ssec:subjective_eval}

The Challenge ranking was determined through a MUSHRA test. To provide a fair test bench, ten signals from the blind test set were handpicked by an expert who only had access to the lossy audio files (zero-filling). In total, two violin excerpts, two cello excerpts, two clarinet excerpts, two double bass excerpts, and two guitar excerpts were selected. All clips are 11.6 seconds long and were presented in full to the assessors. The test was conducted using webMUSHRA \cite{Schoeffler2018webMUSHRAA}, a state-of-the-art Web Audio API-based software compliant to the ITU-R Recommendation BS.1534 \cite{bs1534}.

For each of the ten excerpts, the (undisclosed) clean audio file was used as Reference, whereas the clip degraded with zero-filling was considered as Anchor.
After an initial training page where four audio examples were presented, i.e., two pairs of clean and zero-filling clips, participants were tasked to rate the similarity of each test condition with the Reference on a scale of $0$ to $100$.
On each page, six conditions were assessed, including the output of PARCnet-IS², the hidden Reference, and the Anchor. The names of the test conditions were hidden, and the order of the ten trials, as well as the order of the test items within them, was randomized.
Volume adjustments were only allowed during the training phase. Then, subjects were asked to keep the level constant for the duration of the test. 

A pool of $12$ expert assessors with age ranging from $24$ to $45$ (average: $29.75$), none of whom reported hearing impairments, took part in the listening test. The participants, who completed the test in about $20$ minutes, self-reported an average of $9.6$ years of prior musical training (SD: $5.75$). The assessor pool consisted of members of the Image and Sound Processing Lab (ISPL) at Politecnico di Milano, and had previous experience with MUSHRA tests.

\section{Results}
\begin{figure*}[t]
     \centering
     \subfloat[][MSE]{\includegraphics[width=.32\linewidth]{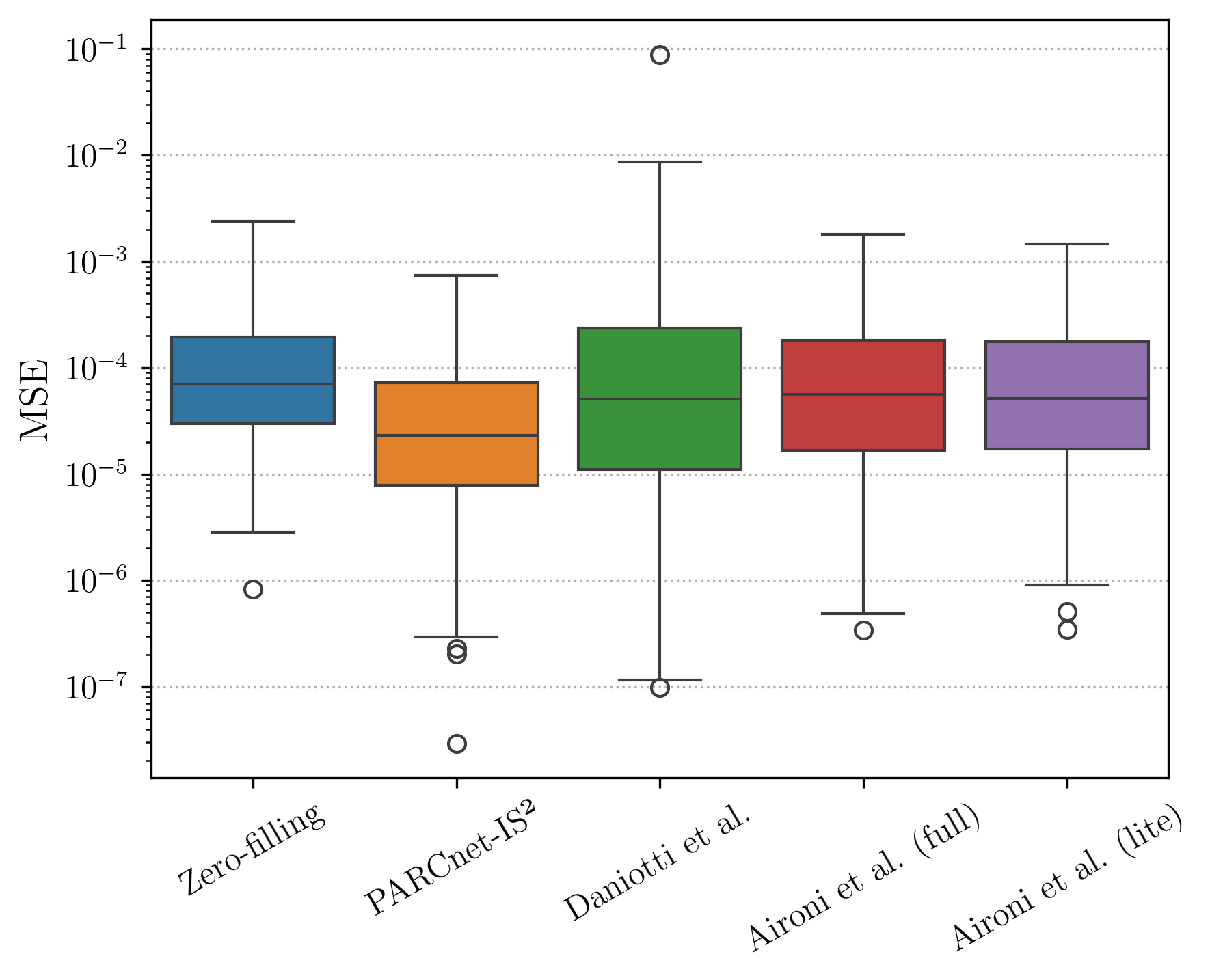}}\label{fig:box_mse}\hspace{2cm}
     \subfloat[][SDR]{\includegraphics[width=.32\linewidth]{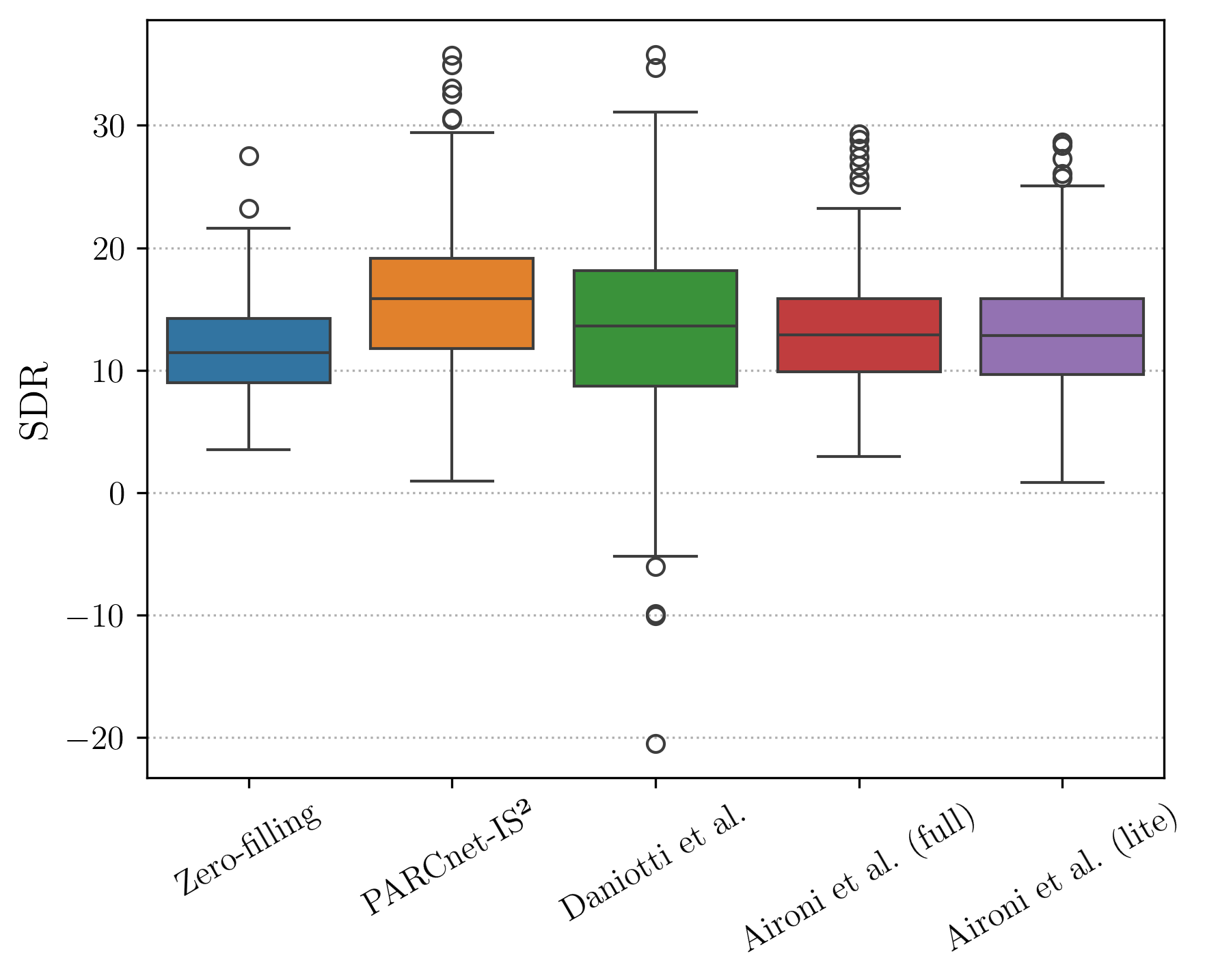}\label{fig:box_sdr}}\\
     \subfloat[][SI-SDR]{\includegraphics[width=.32\linewidth]{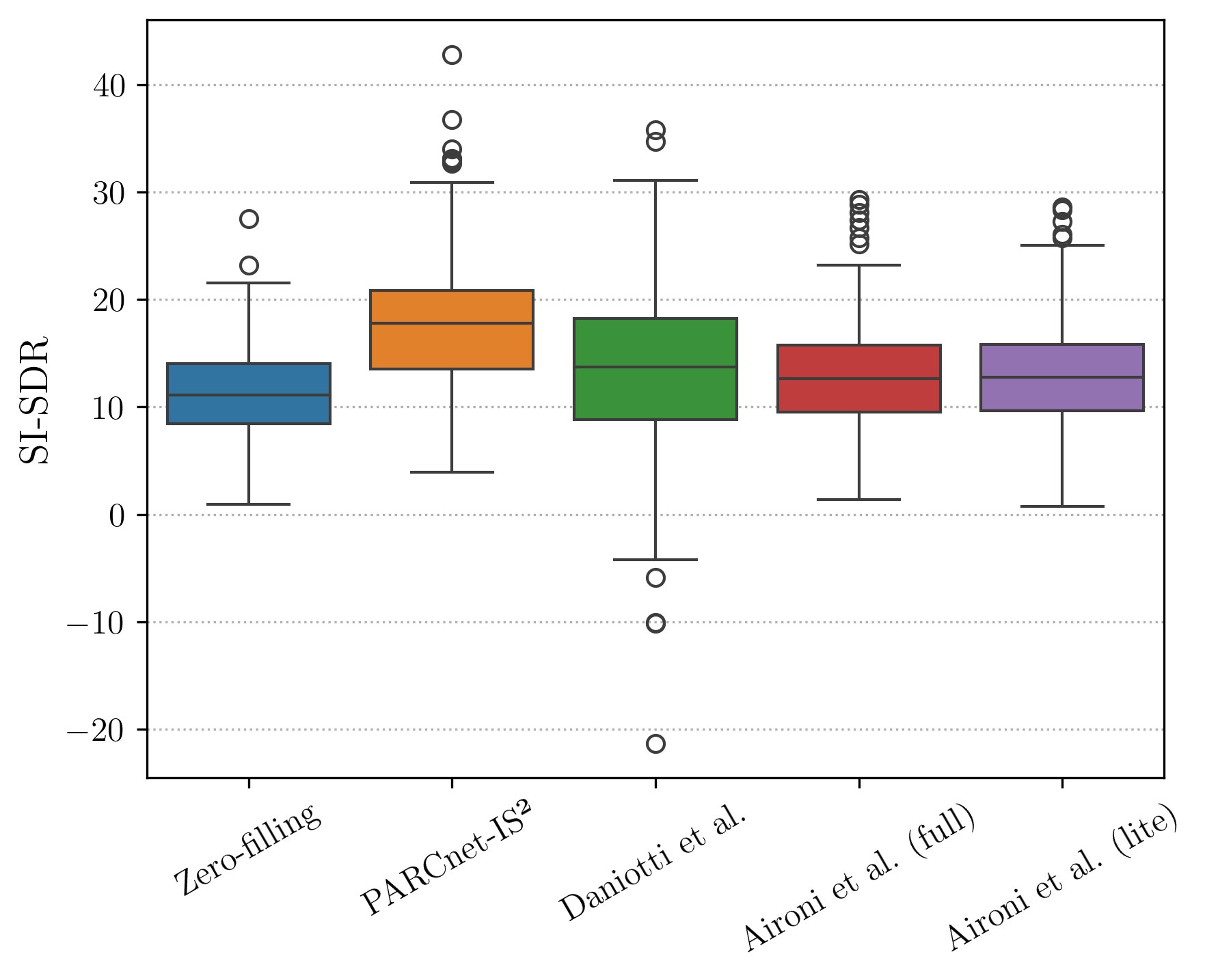}\label{fig:box_si-sdr}}\hspace{2cm}
     \subfloat[][LSD]{\includegraphics[width=.32\linewidth]{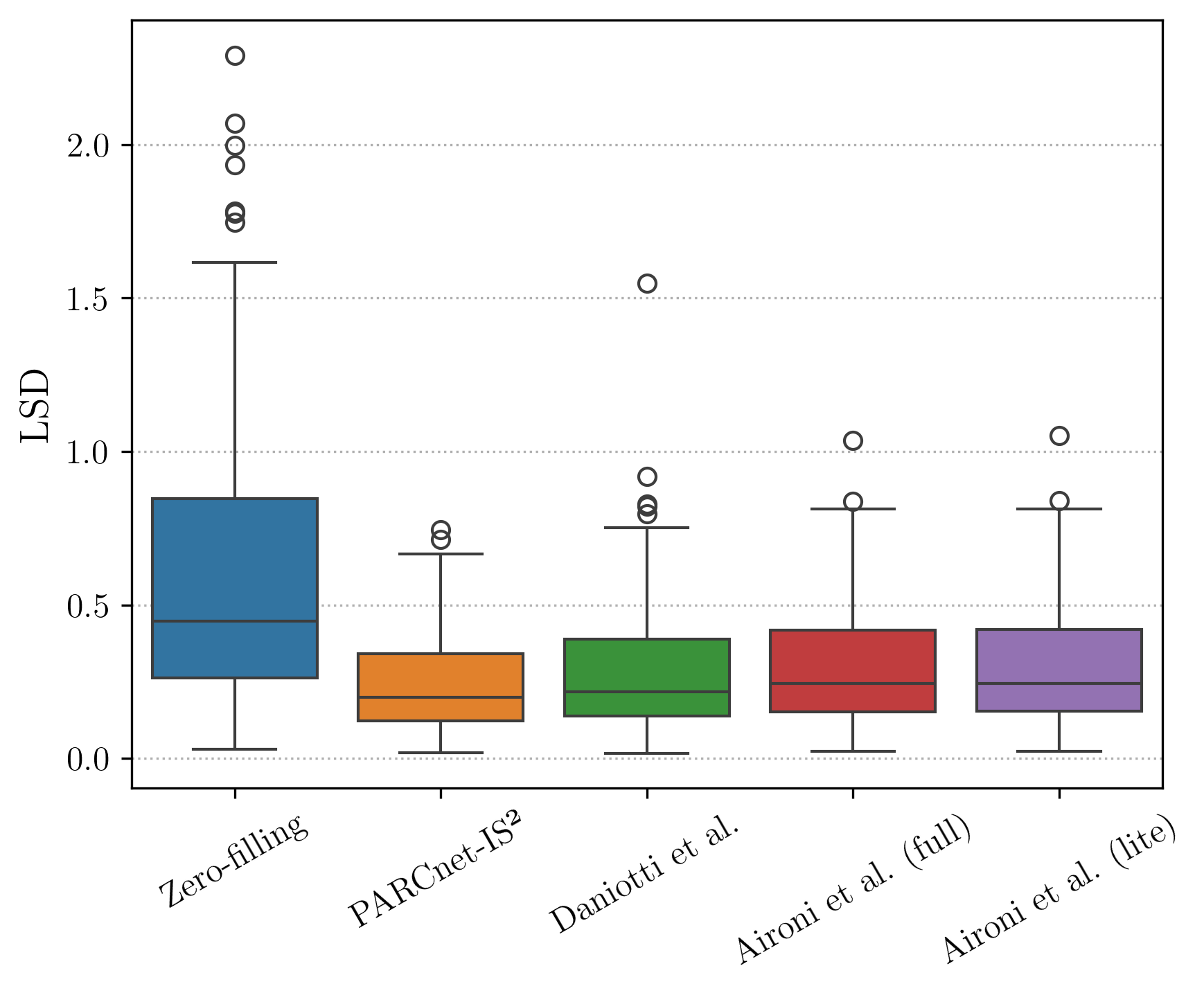}\label{fig:box_lsd}}\\
     \subfloat[][MCD]{\includegraphics[width=.32\linewidth]{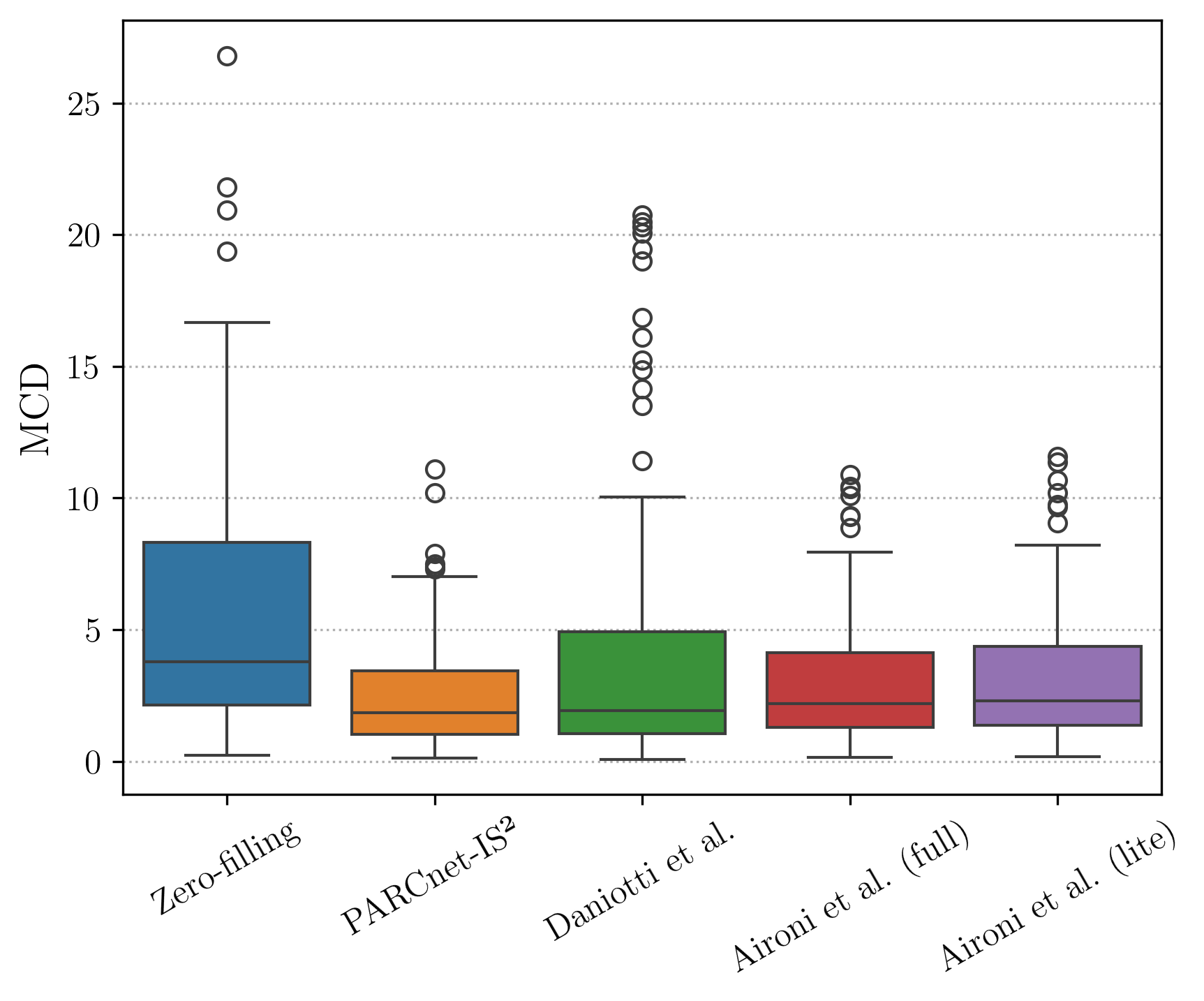}}\label{fig:box_mcd}\hspace{2cm}
     \subfloat[][PEAQ]{\includegraphics[width=.32\linewidth]{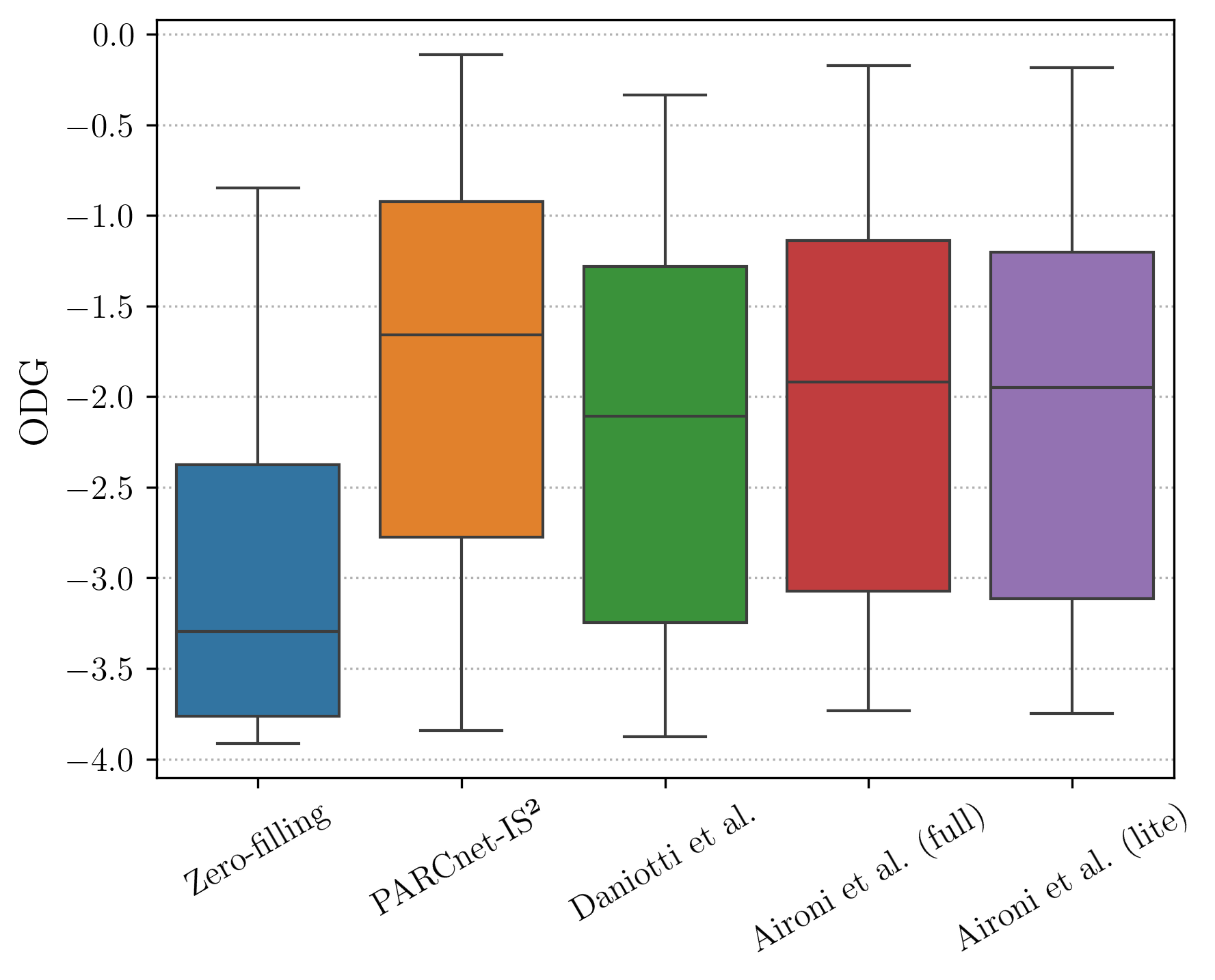}\label{fig:box_peaq}}\\
     \subfloat[][PLCMOS]{\includegraphics[width=.32\linewidth]{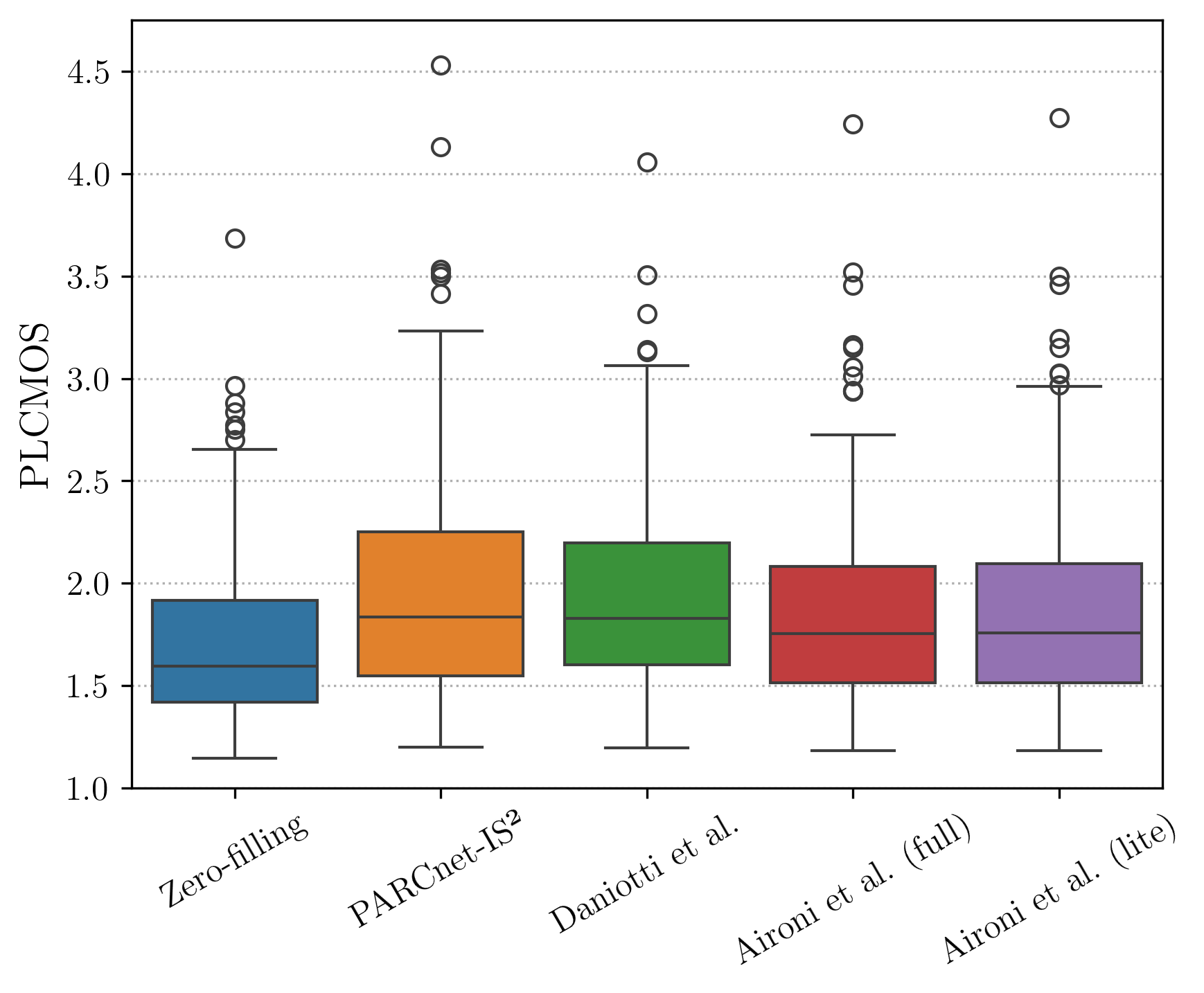}\label{fig:box_plcmos}}
     \caption{Objective metrics computed on the entire blind test set.}
     \label{fig:objective_metrics_boxplot}
\end{figure*}

Figure~\ref{fig:ranking} shows the average scores and $95\%$ confidence intervals obtained across all trials in the MUSHRA test (top) and the number of trials won by each PLC method (bottom).
Table~\ref{tab:objective_metrics} reports the average objective metrics outlined in Section~\ref{ssec:objective_eval}. Figure~\ref{fig:objective_metrics_boxplot} depicts the box-and-whisker plots for each metric, whereas Figures~\ref{fig:mushra_trials_boxplot} and \ref{fig:mushra_trials_barplot} shows the box-and-whisker plots and the average scores of every trial in the listening test, respectively.  

Table~\ref{tab:objective_metrics} indicates that the baseline method, on average, outperforms the submitted PLC systems across all objective metrics. 
%Notably, PEAQ indicates PARCnet-IS² as the only method to yield results that lie between \textit{Slightly annoying} and \textit{Perceptible, but not annoying} (see Table~\ref{tab:impairment_description}). 
%
These results appear to be confirmed by the outcome of the MUSHRA test in Figure~\ref{fig:ranking}, where PARCnet-IS² won $9$ out of $10$ trials and Aironi et al.\ (full) won one trial (Violin~\#2). This led to the final ranking given in Table~\ref{tab:ranking}.

%\section{Conclusions}

\section*{Acknowledgment}
The organizers wish to thank Luca Turchet and the Organizing Committee of the 2\textsuperscript{nd} IEEE International Workshop on Networked Immersive Audio.

\bibliographystyle{IEEEtran}
\bibliography{bibliography}

\begin{figure*}[t]
     \centering
     \subfloat[][Cello \#1]{\includegraphics[width=.32\linewidth]{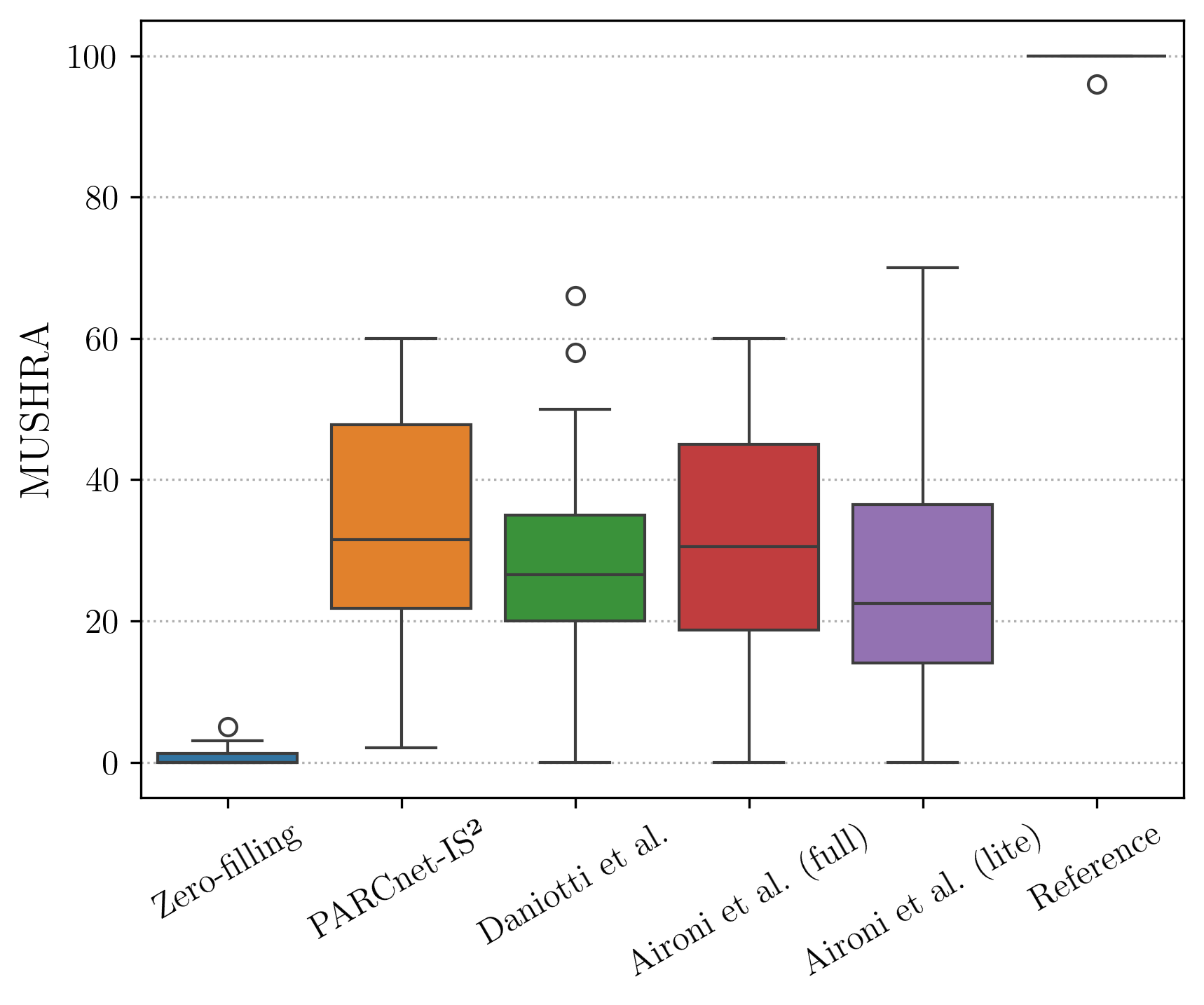}\label{fig:cello1}}\hspace{2cm}
     \subfloat[][Cello \#2]{\includegraphics[width=.32\linewidth]{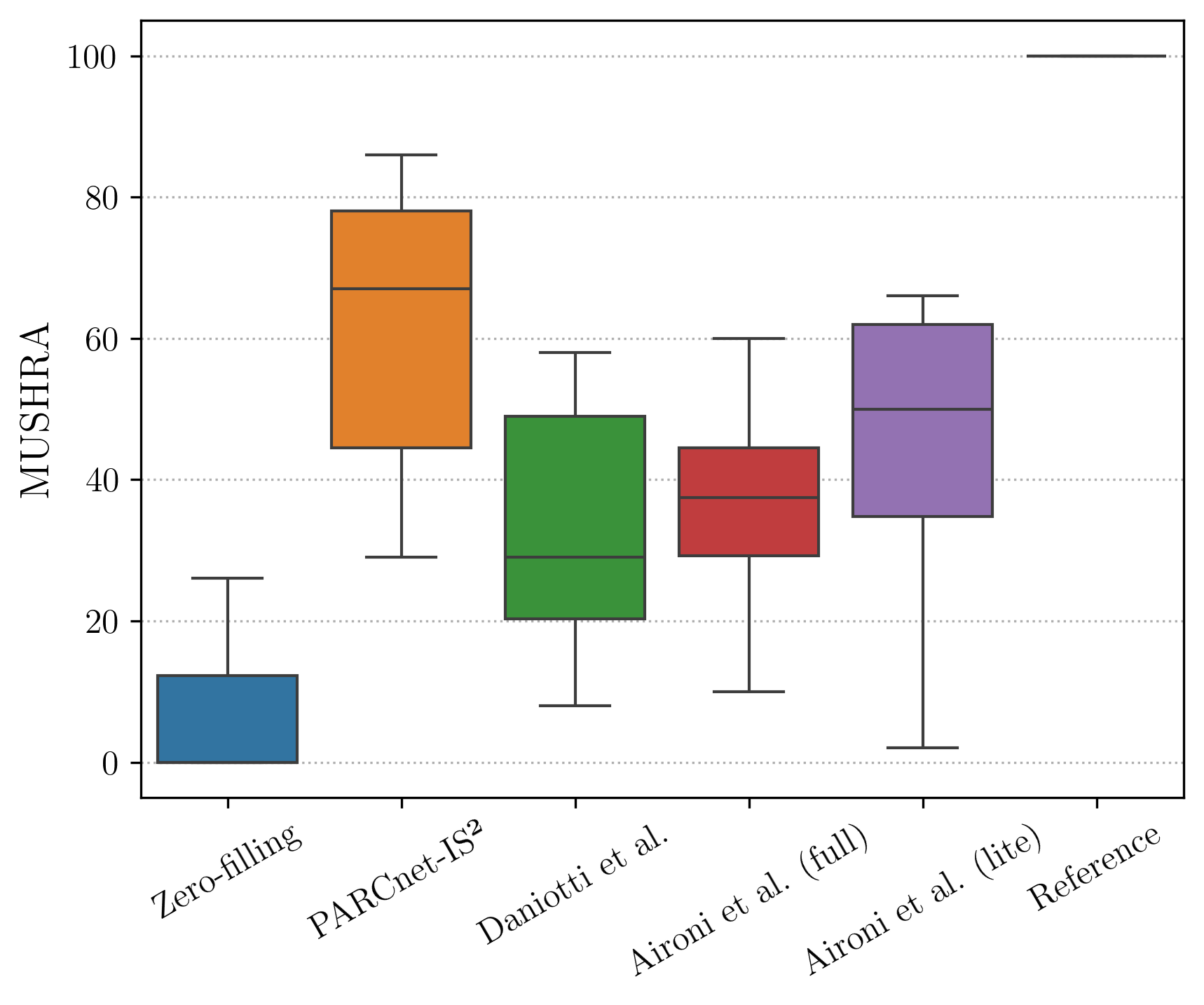}\label{fig:cello2}}\\
     \subfloat[][Clarinet \#1]{\includegraphics[width=.32\linewidth]{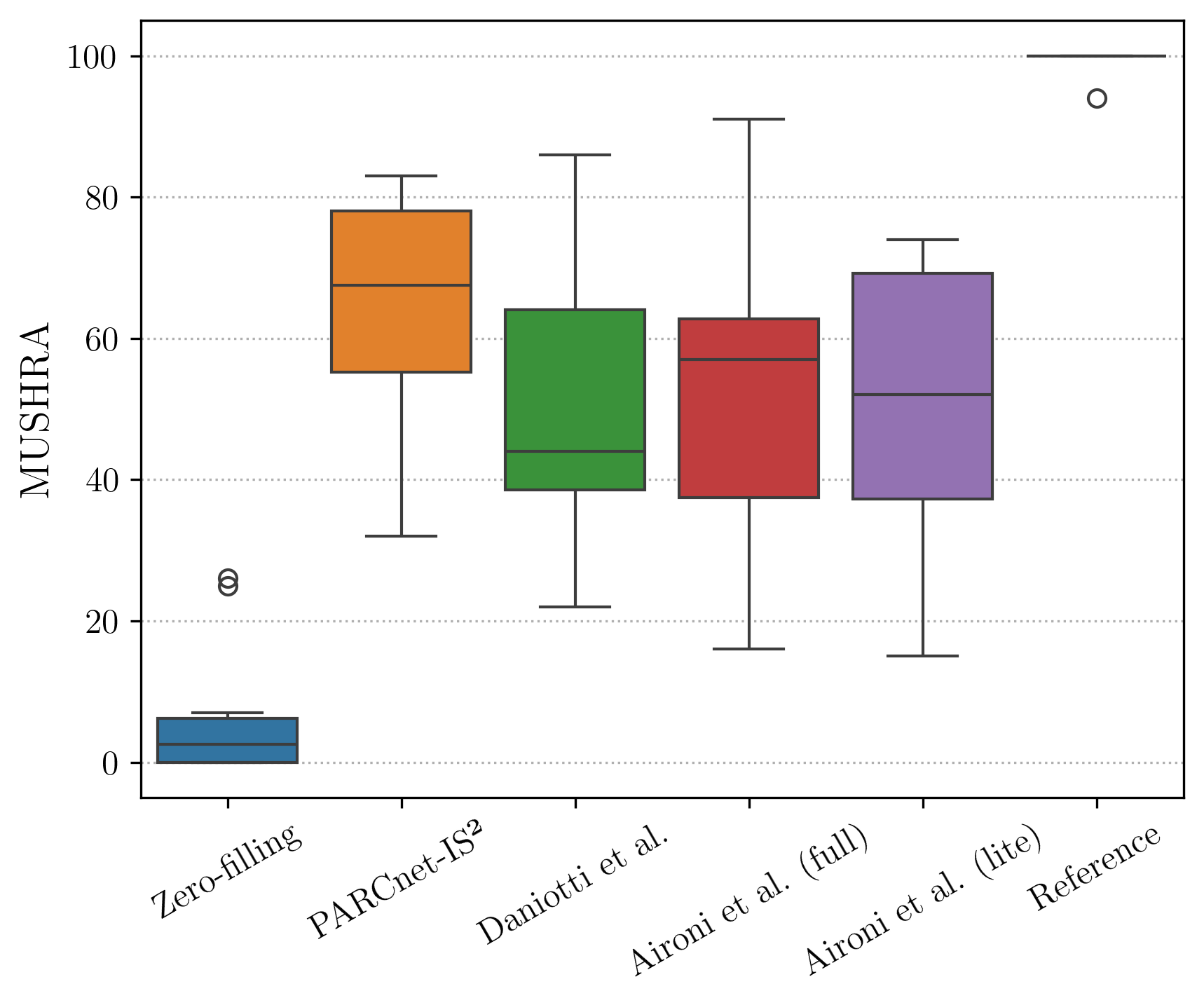}\label{fig:clarinet1}}\hspace{2cm}
     \subfloat[][Clarinet \#2]{\includegraphics[width=.32\linewidth]{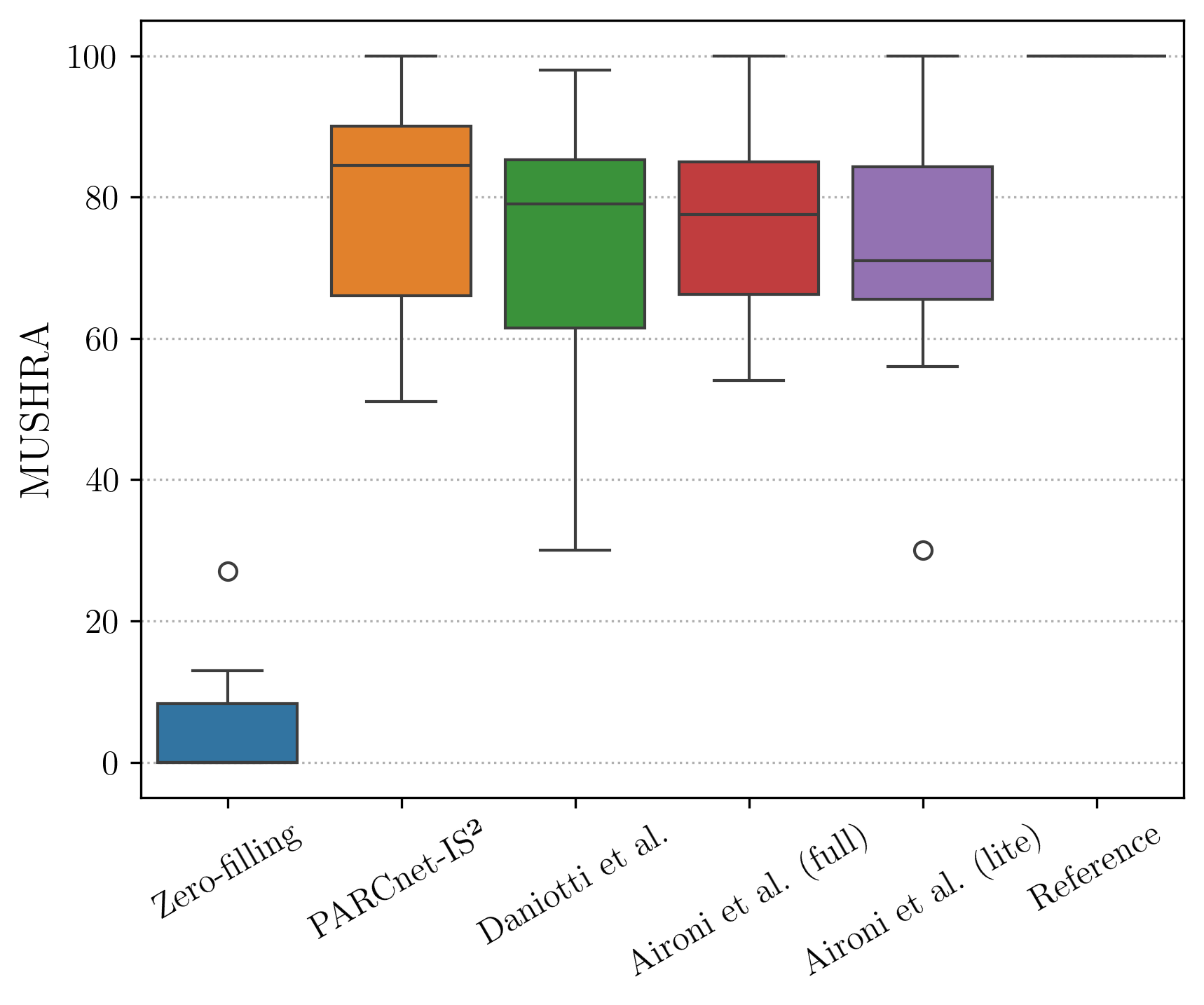}\label{fig:clarinet2}}\\
     \subfloat[][Double Bass \#1]{\includegraphics[width=.32\linewidth]{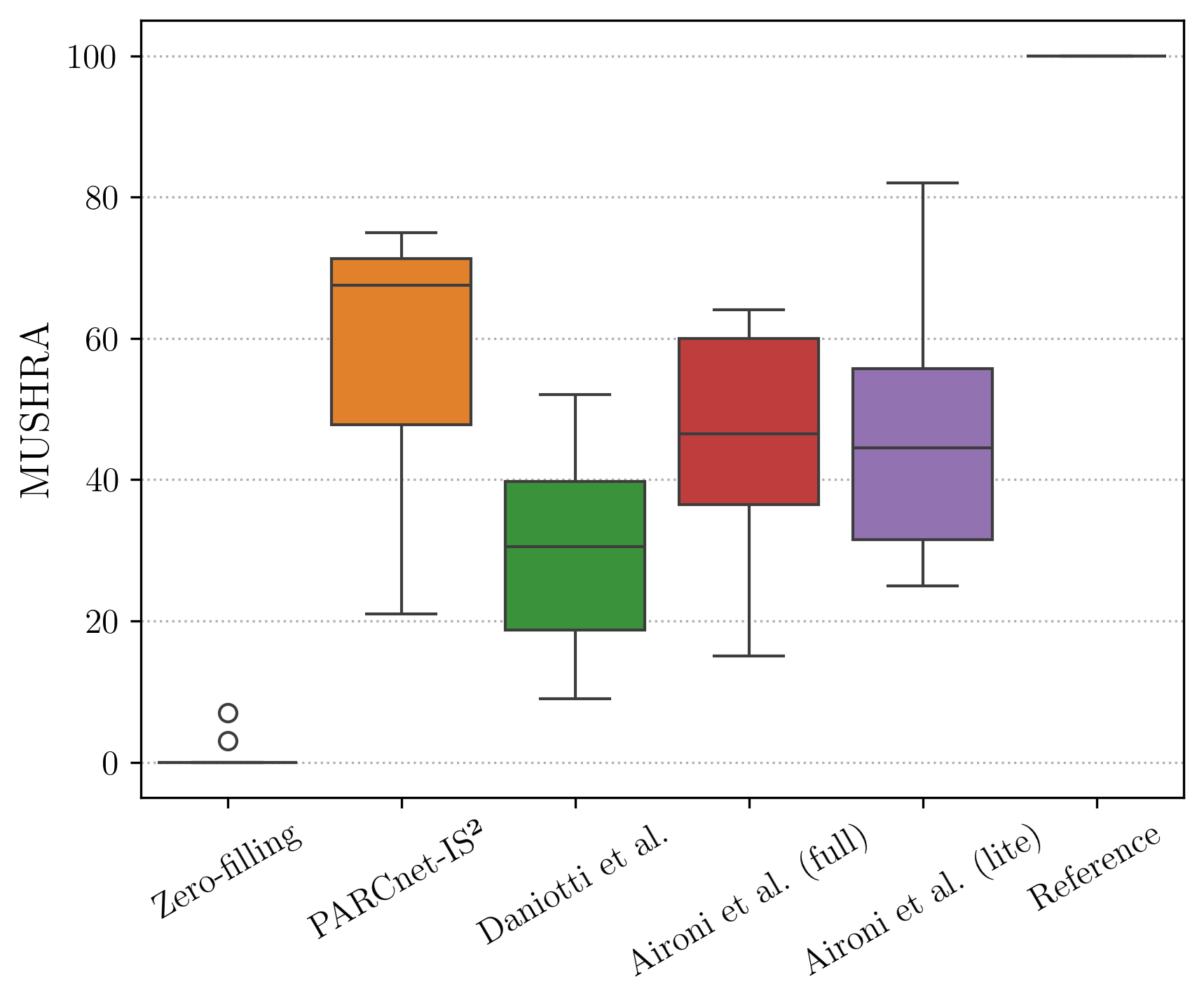}\label{fig:doublebass1}}\hspace{2cm}
     \subfloat[][Double Bass \#2]{\includegraphics[width=.32\linewidth]{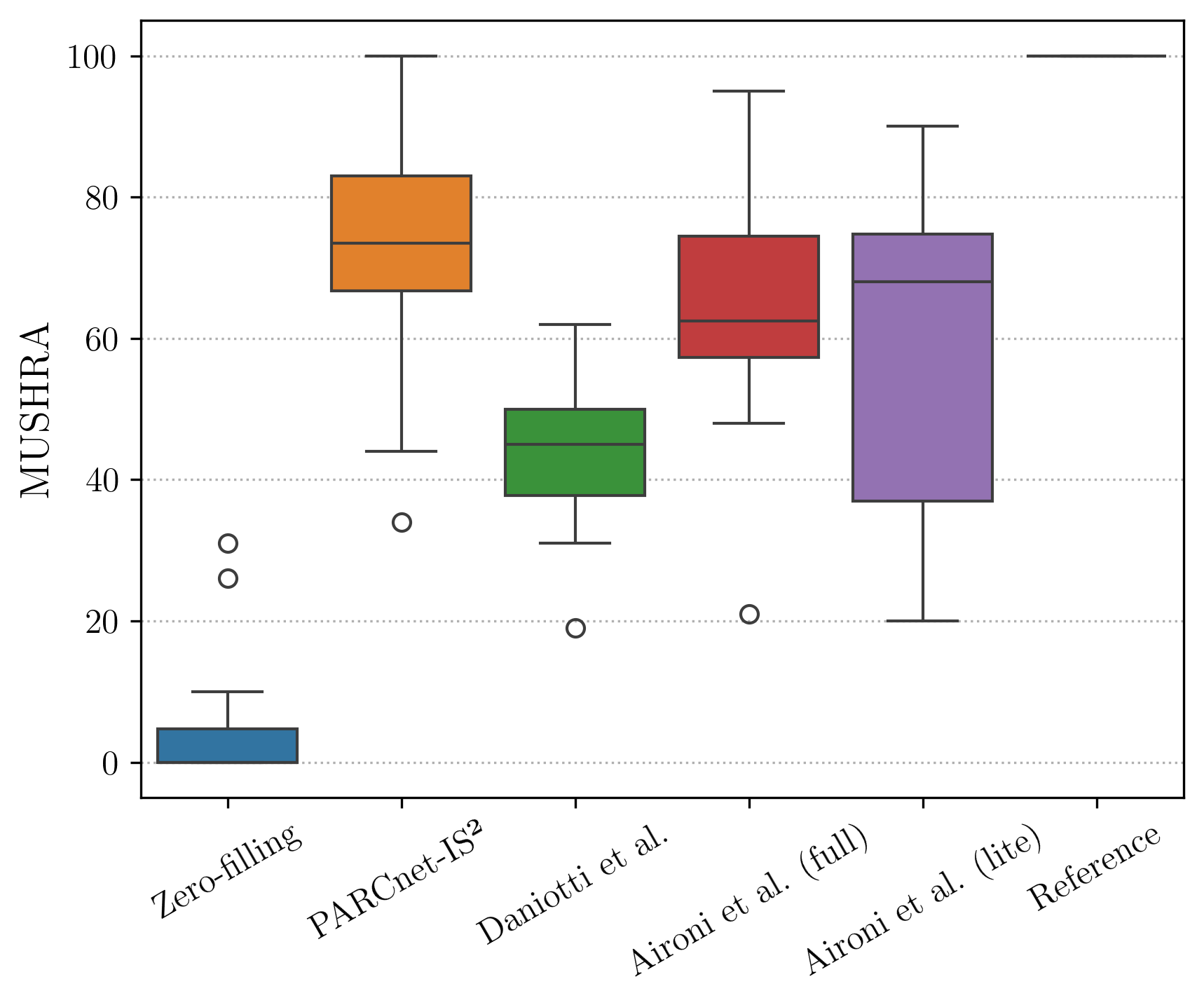}\label{fig:doublebass2}}\\
      \subfloat[][Guitar \#1]{\includegraphics[width=.32\linewidth]{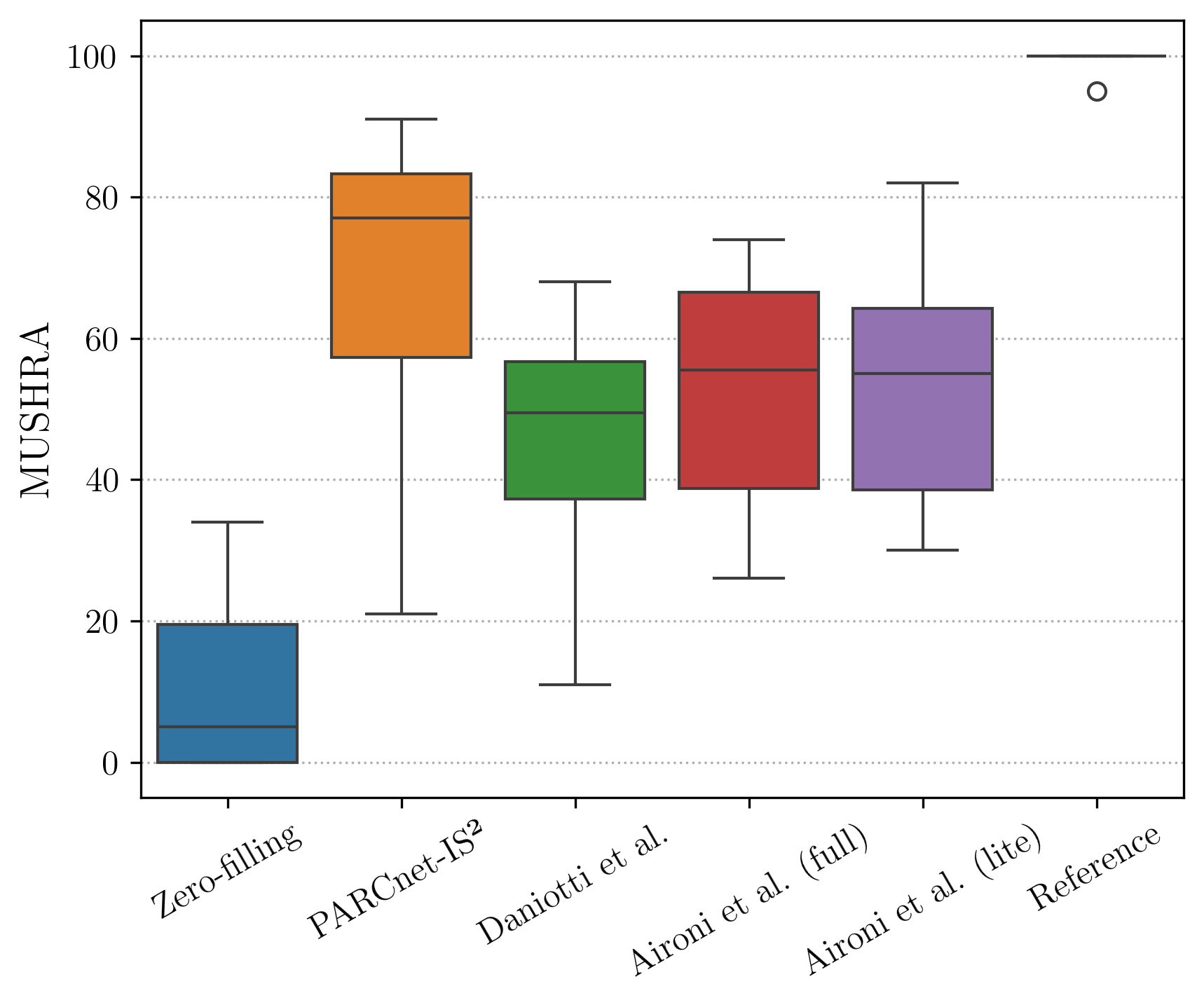}\label{fig:guitar1}}\hspace{2cm}
     \subfloat[][Guitar \#2]{\includegraphics[width=.32\linewidth]{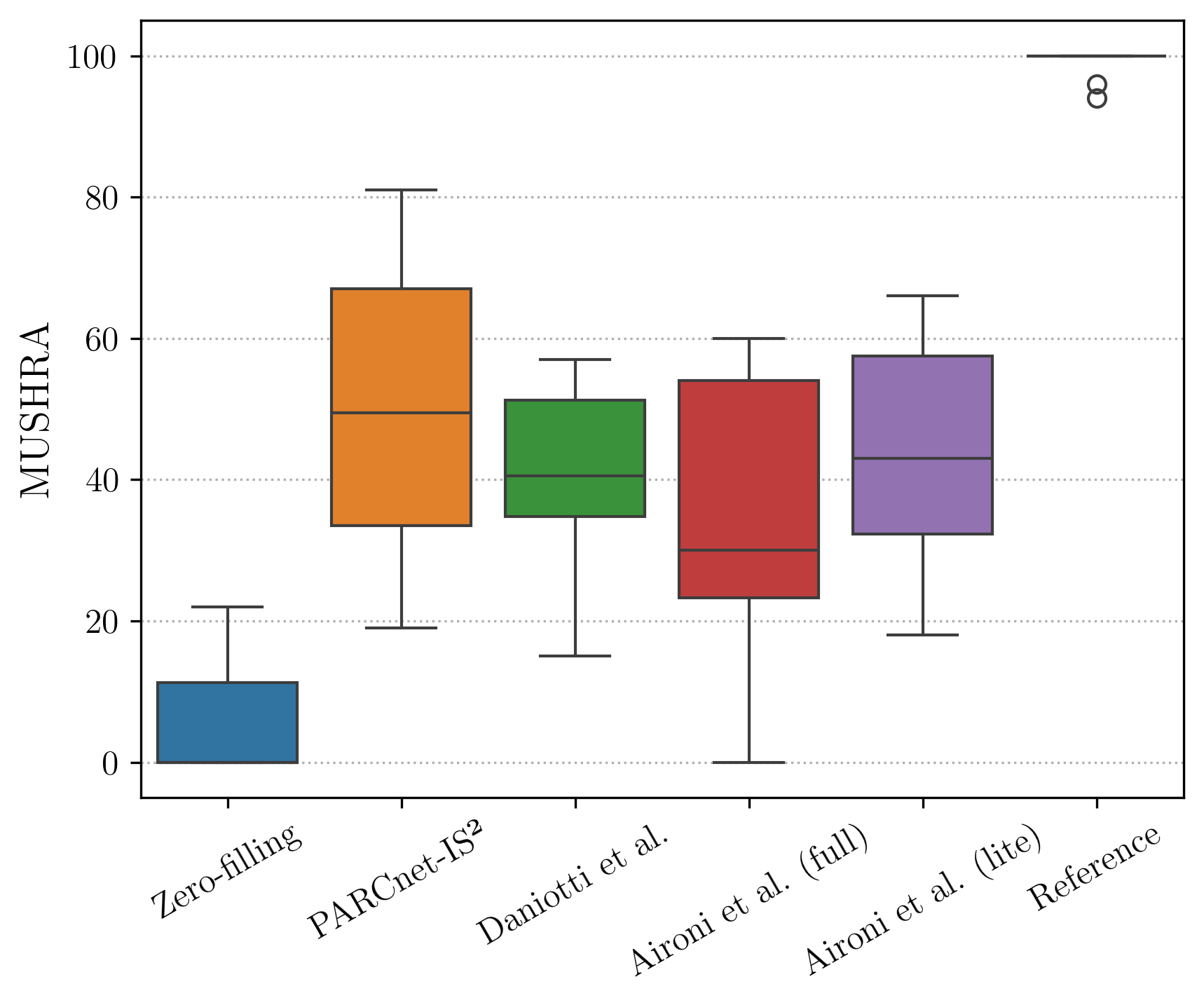}\label{fig:guitar2}}\\
    \caption{Box-and-whisker plots of the individual trials in the MUSHRA test.}
     \label{fig:mushra_trials_boxplot}
\end{figure*}
\begin{figure*}[t!]\ContinuedFloat
    \centering
    \subfloat[][Violin \#1]{\includegraphics[width=.32\linewidth]{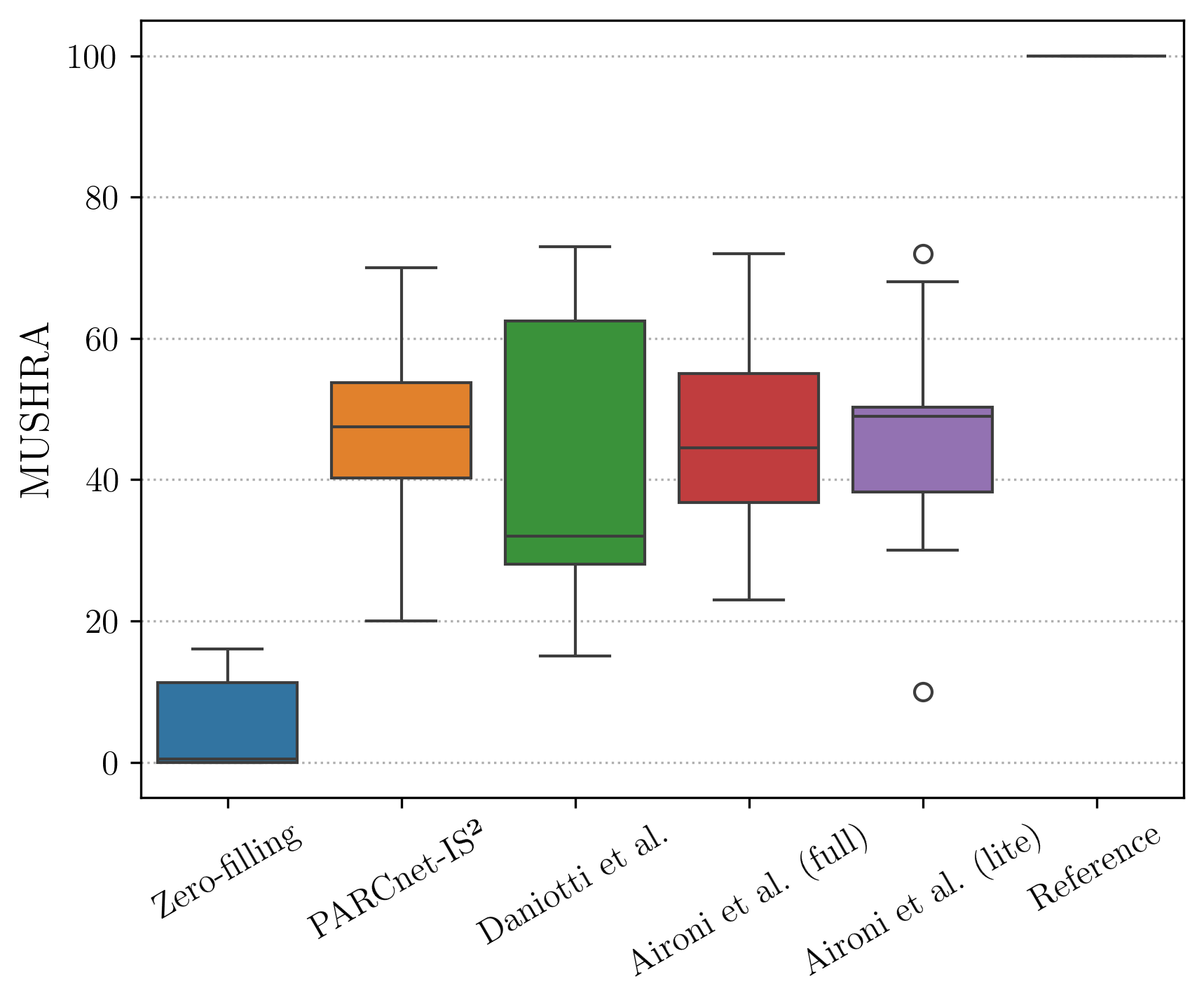}\label{fig:violin1}}\hspace{2cm}
     \subfloat[][Violin \#2]{\includegraphics[width=.32\linewidth]{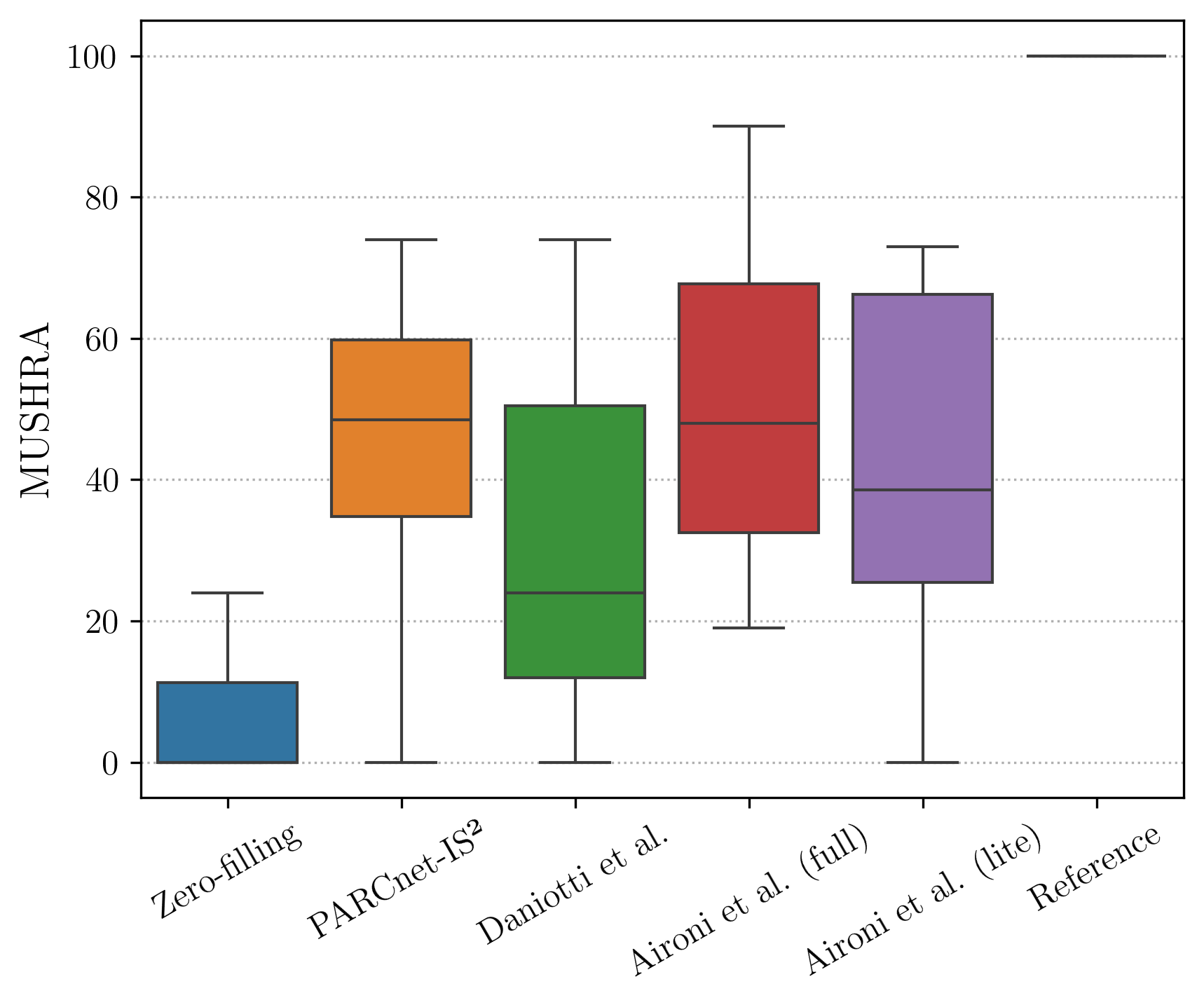}\label{fig:violin2}}\\
    \caption{Box-and-whisker plots of the individual trials in the MUSHRA test (cont.)}
\end{figure*}

\begin{figure*}[t]
     \centering
     \subfloat[][Cello \#1]{\includegraphics[width=.32\linewidth]{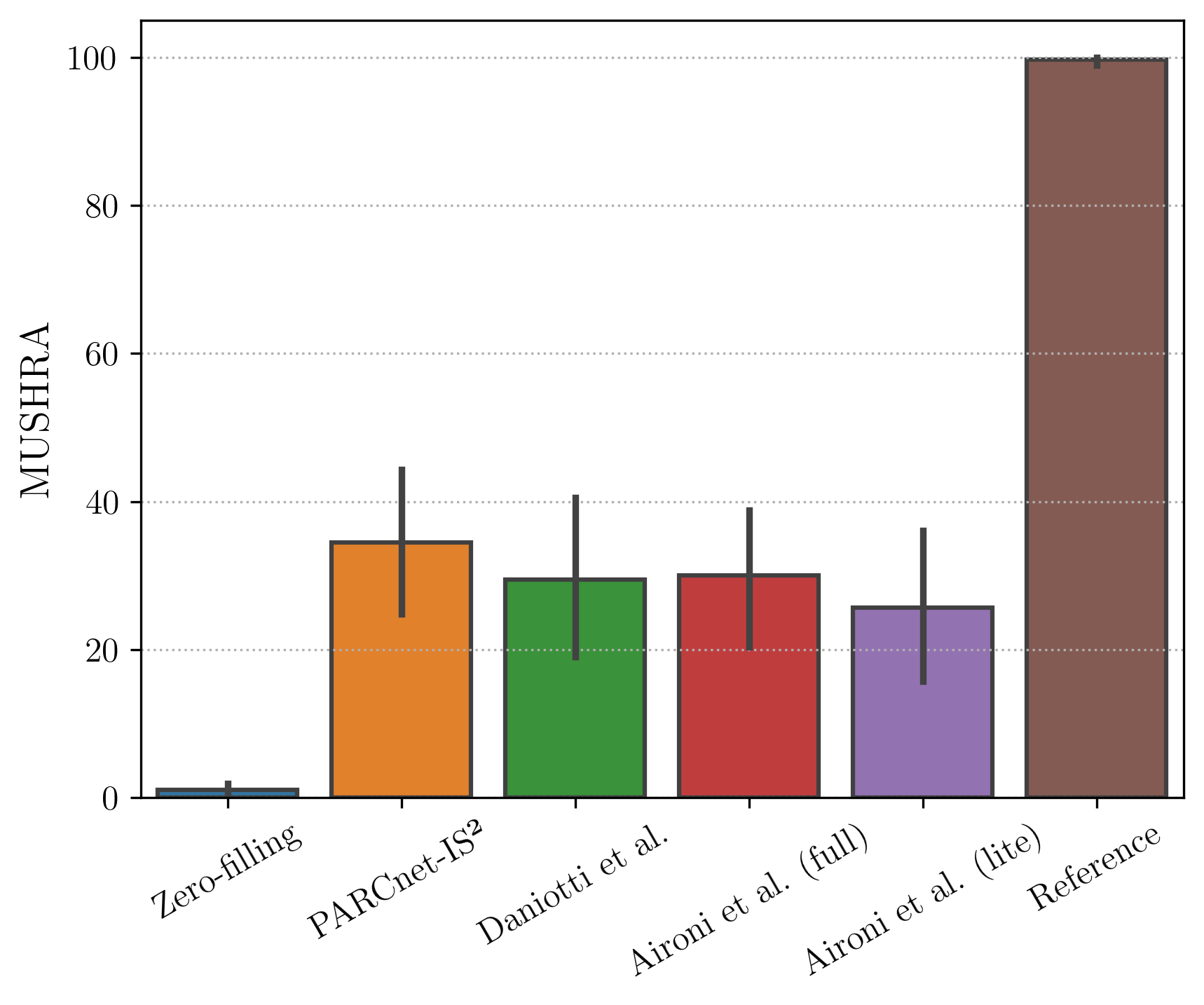}\label{fig:bar-bar-cello1}}\hspace{2cm}
     \subfloat[][Cello \#2]{\includegraphics[width=.32\linewidth]{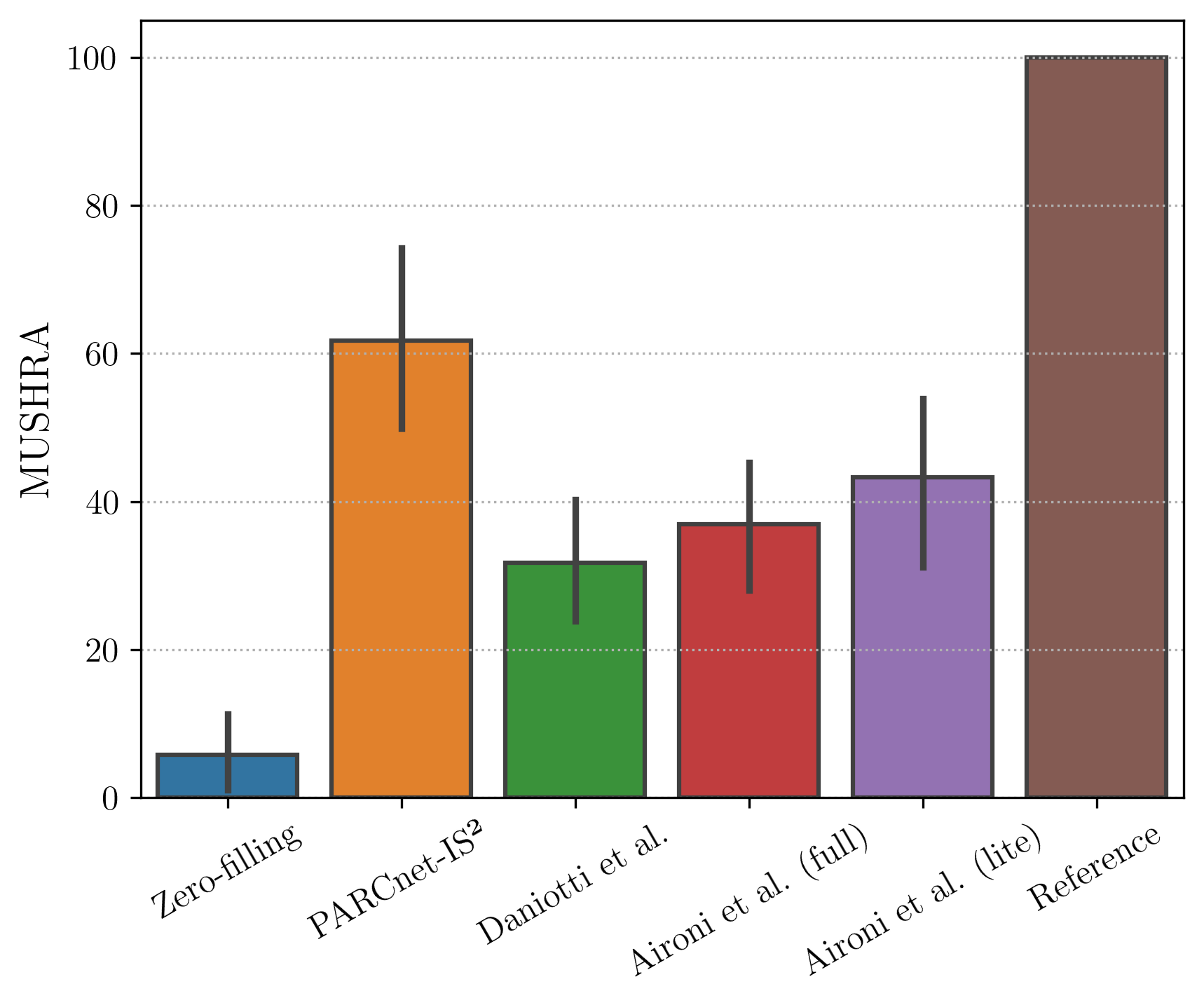}\label{fig:bar-bar-cello2}}\\
     \subfloat[][Clarinet \#1]{\includegraphics[width=.32\linewidth]{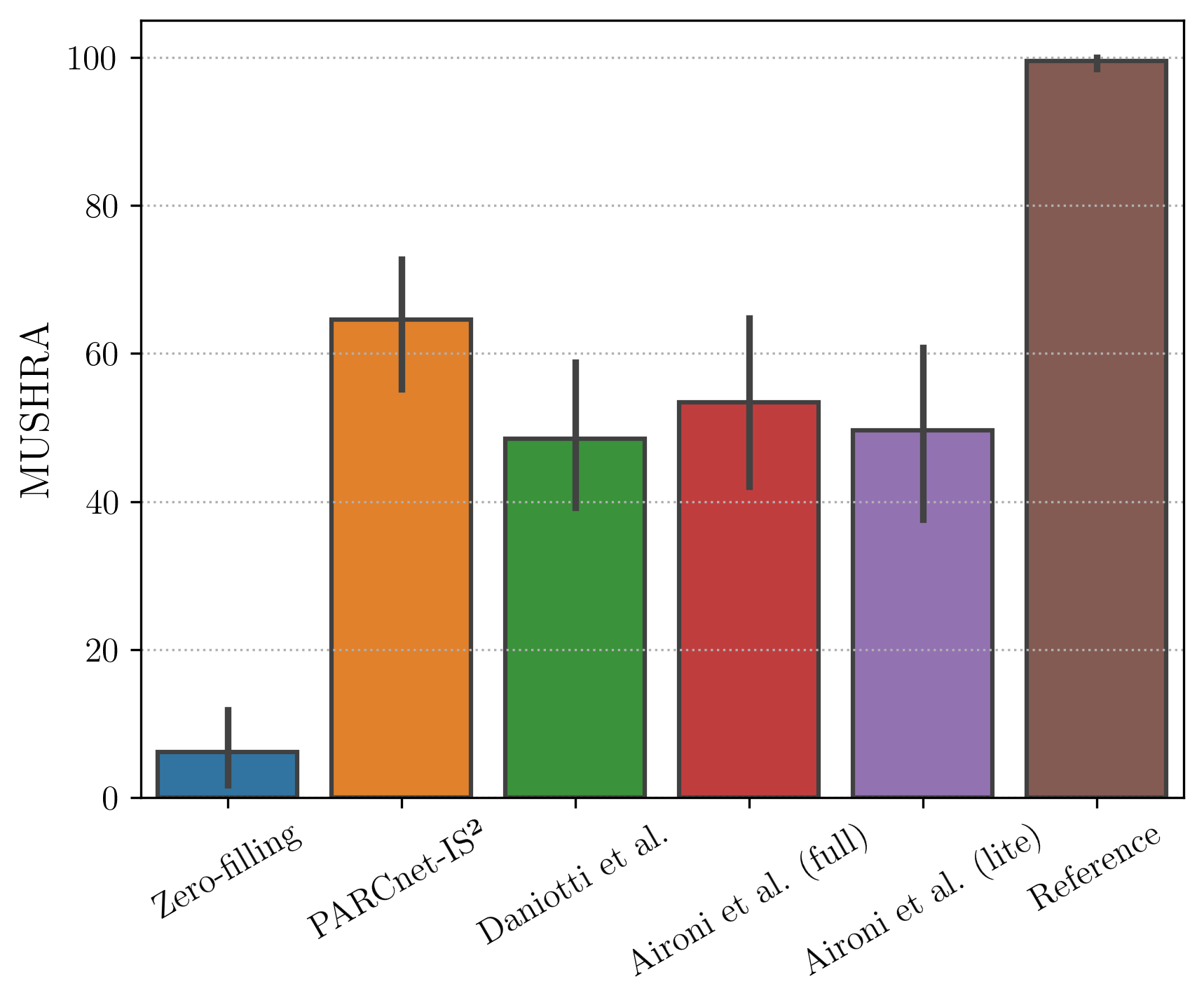}\label{fig:bar-bar-clarinet1}}\hspace{2cm}
     \subfloat[][Clarinet \#2]{\includegraphics[width=.32\linewidth]{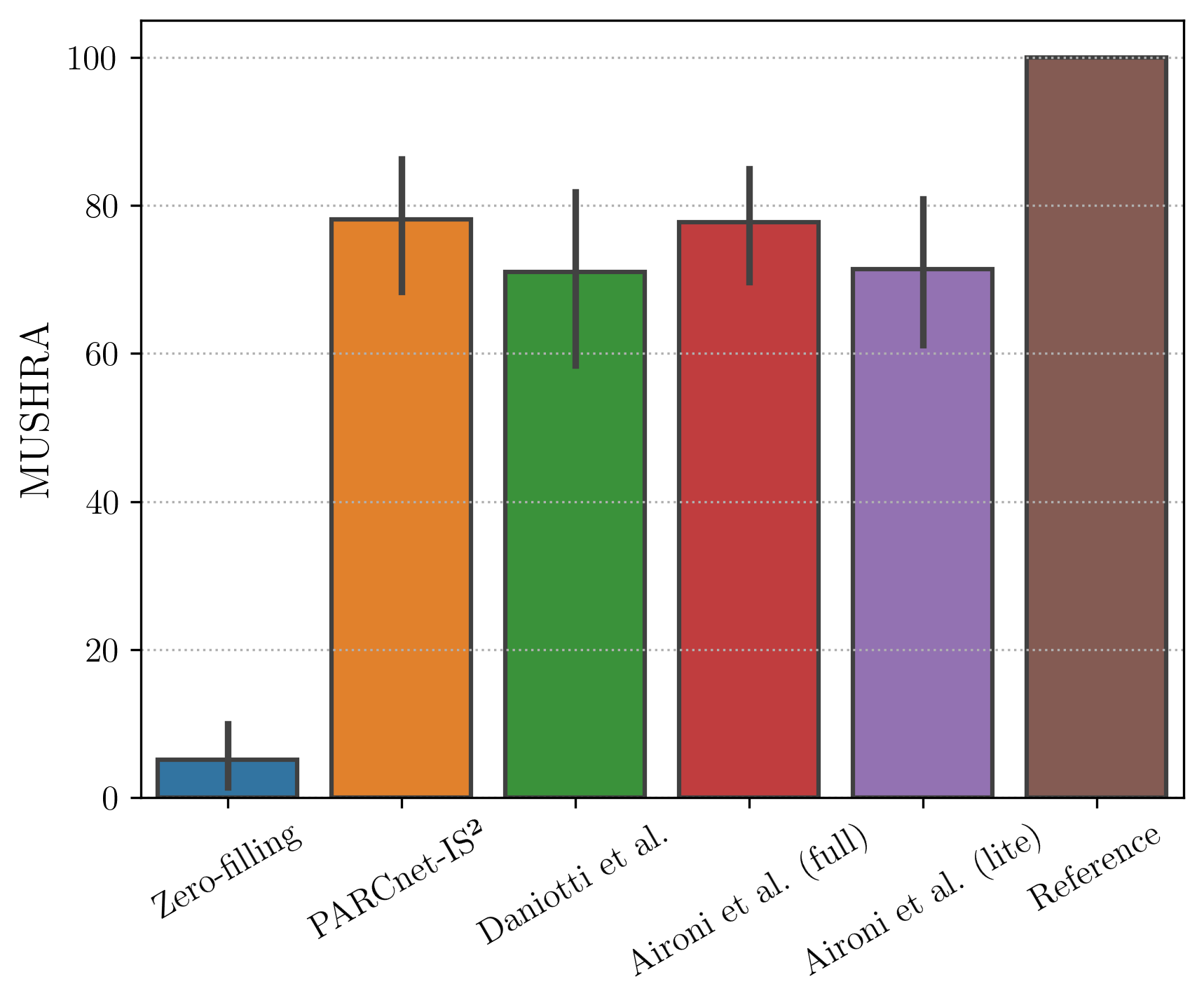}\label{fig:bar-bar-clarinet2}}\\
    \caption{Average scores and $95\%$ confidence intervals of the individual trials in the MUSHRA test.}
     \label{fig:mushra_trials_barplot}
\end{figure*}
\begin{figure*}[t!]\ContinuedFloat
    \centering
    \subfloat[][Double Bass \#1]{\includegraphics[width=.32\linewidth]{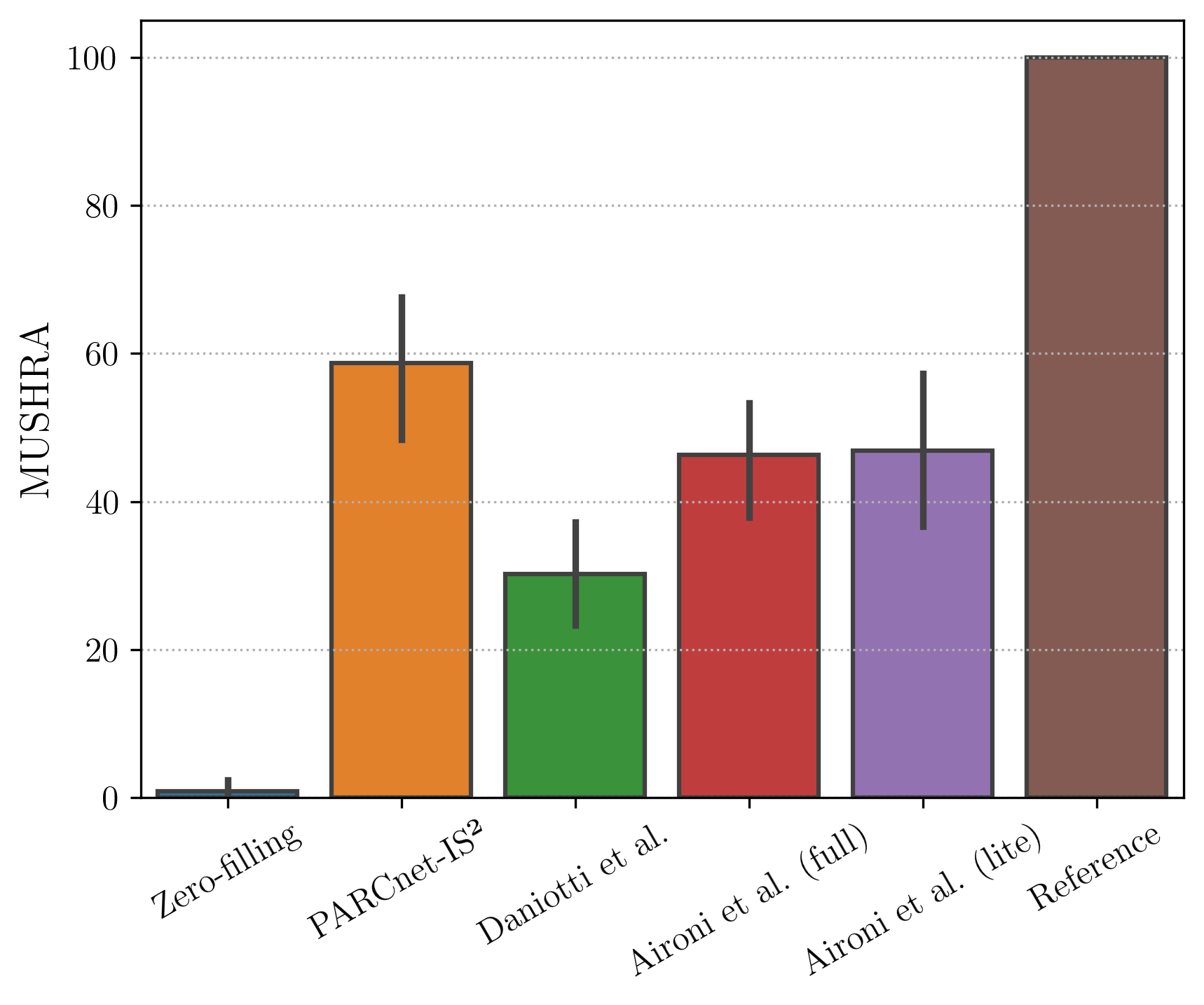}\label{fig:bar-doublebass1}}\hspace{2cm}
     \subfloat[][Double Bass \#2]{\includegraphics[width=.32\linewidth]{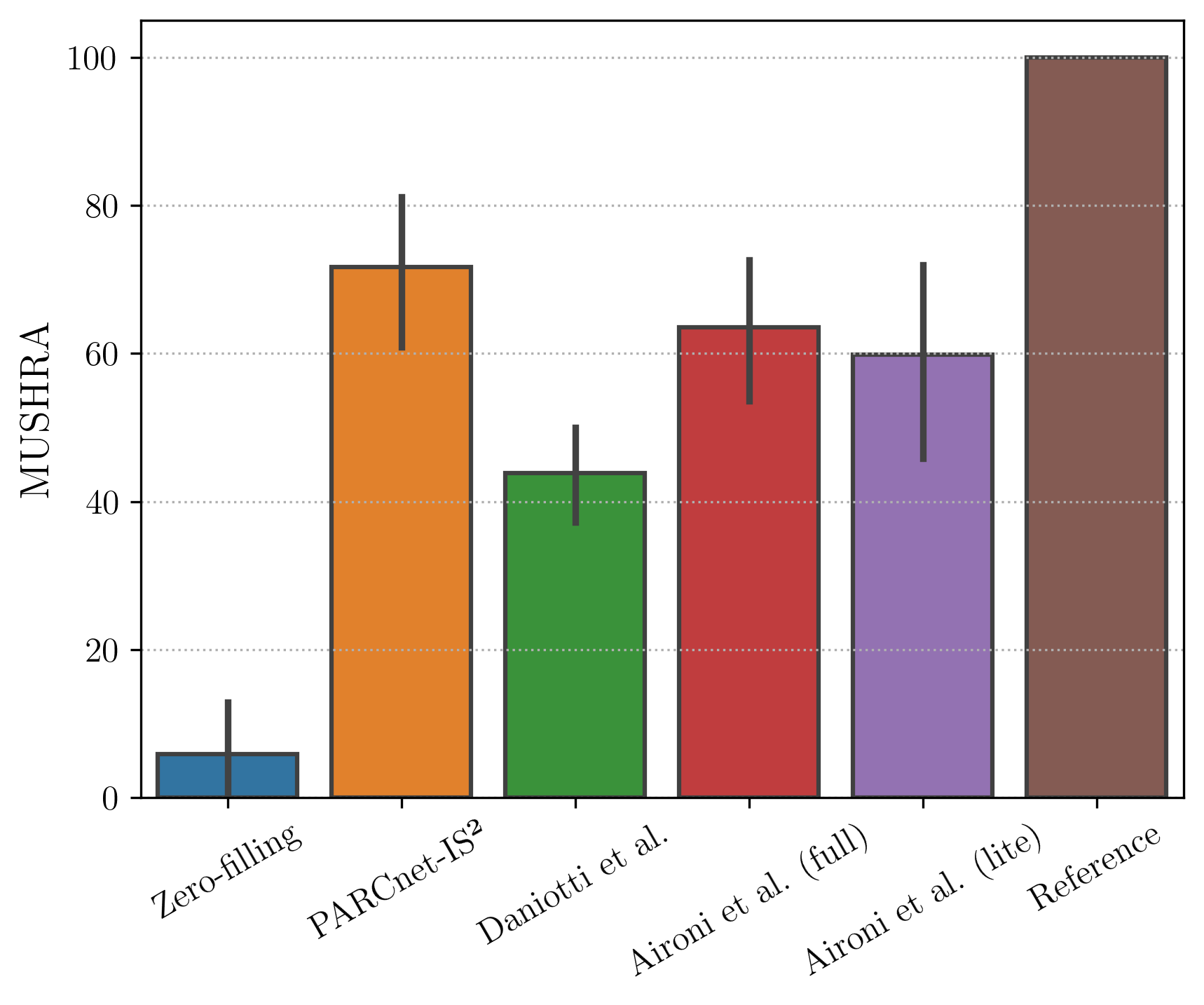}\label{fig:bar-doublebass2}}\\
    \subfloat[][Guitar \#1]{\includegraphics[width=.32\linewidth]{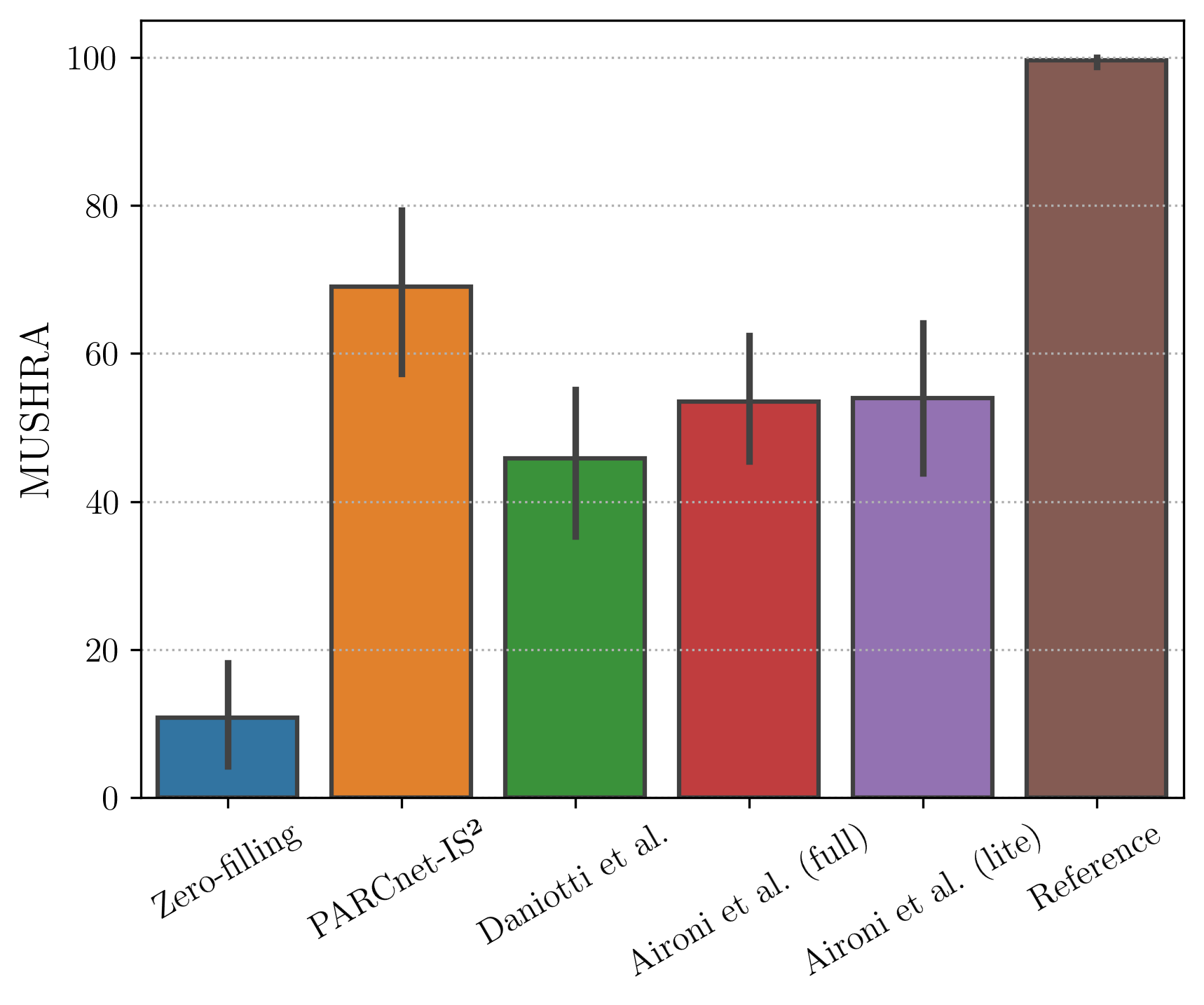}\label{fig:bar-guitar1}}\hspace{2cm}
     \subfloat[][Guitar \#2]{\includegraphics[width=.32\linewidth]{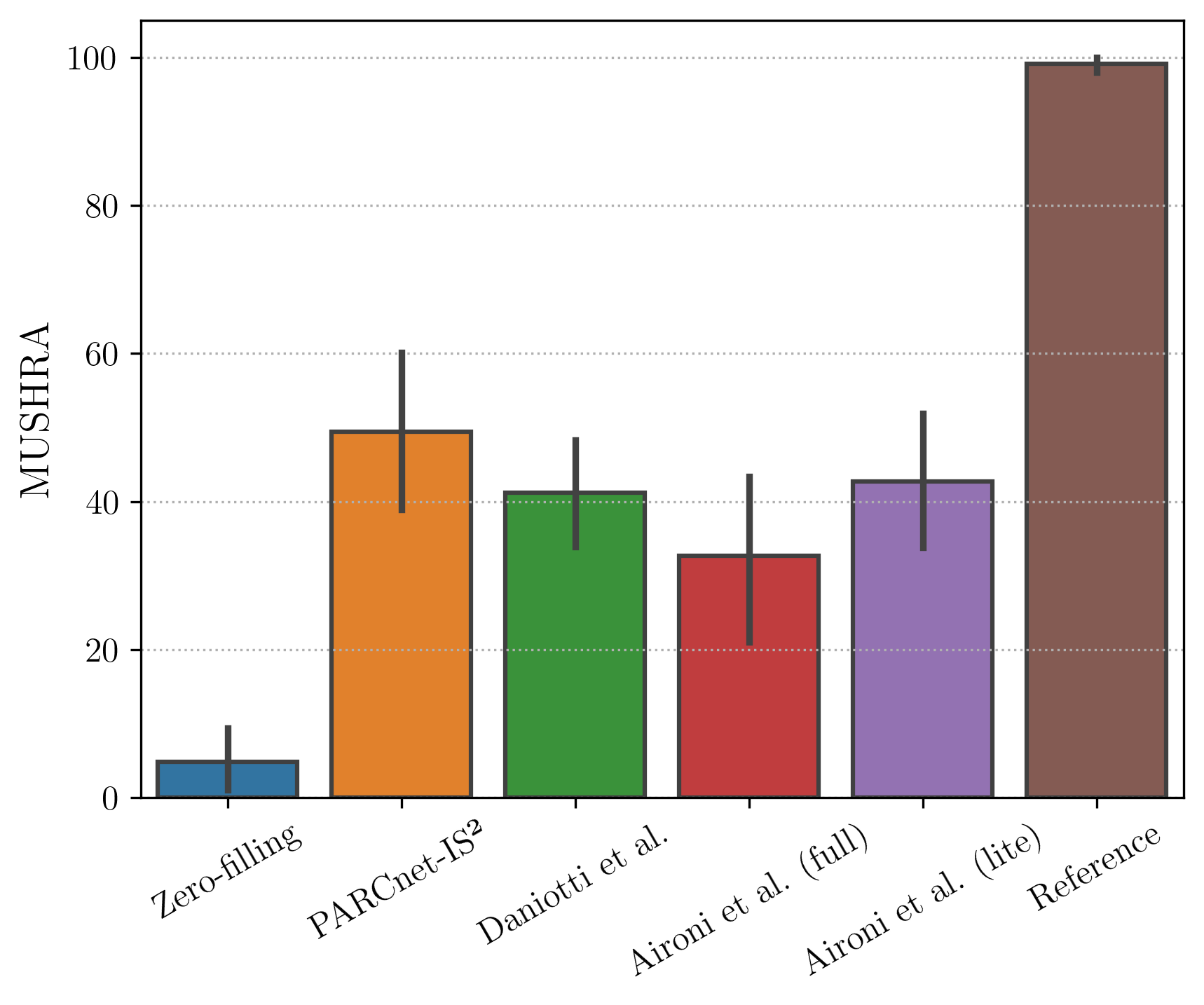}\label{fig:bar-guitar2}}\\
    \subfloat[][Violin \#1]{\includegraphics[width=.32\linewidth]{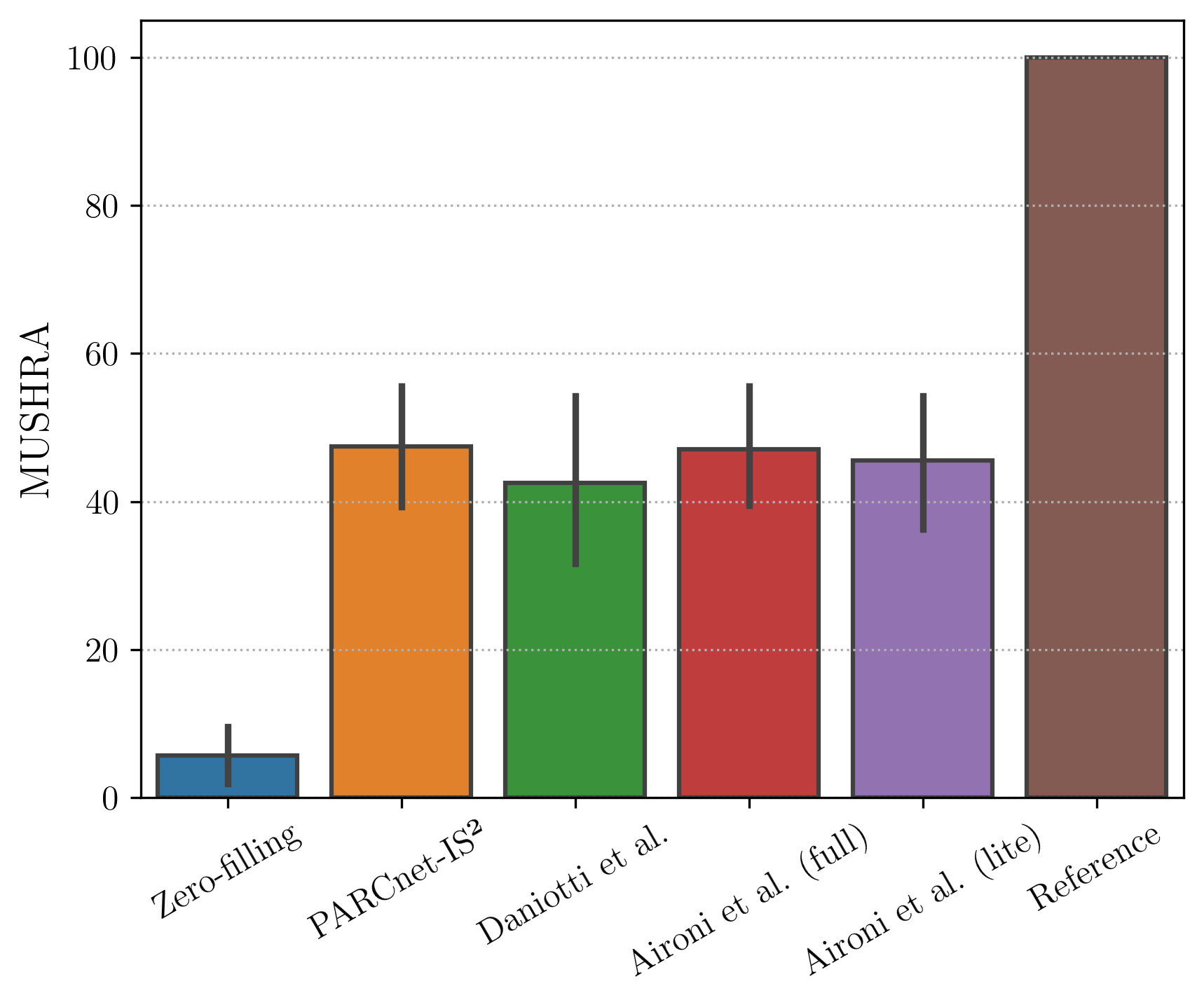}\label{fig:bar-violin1}}\hspace{2cm}
     \subfloat[][Violin \#2]{\includegraphics[width=.32\linewidth]{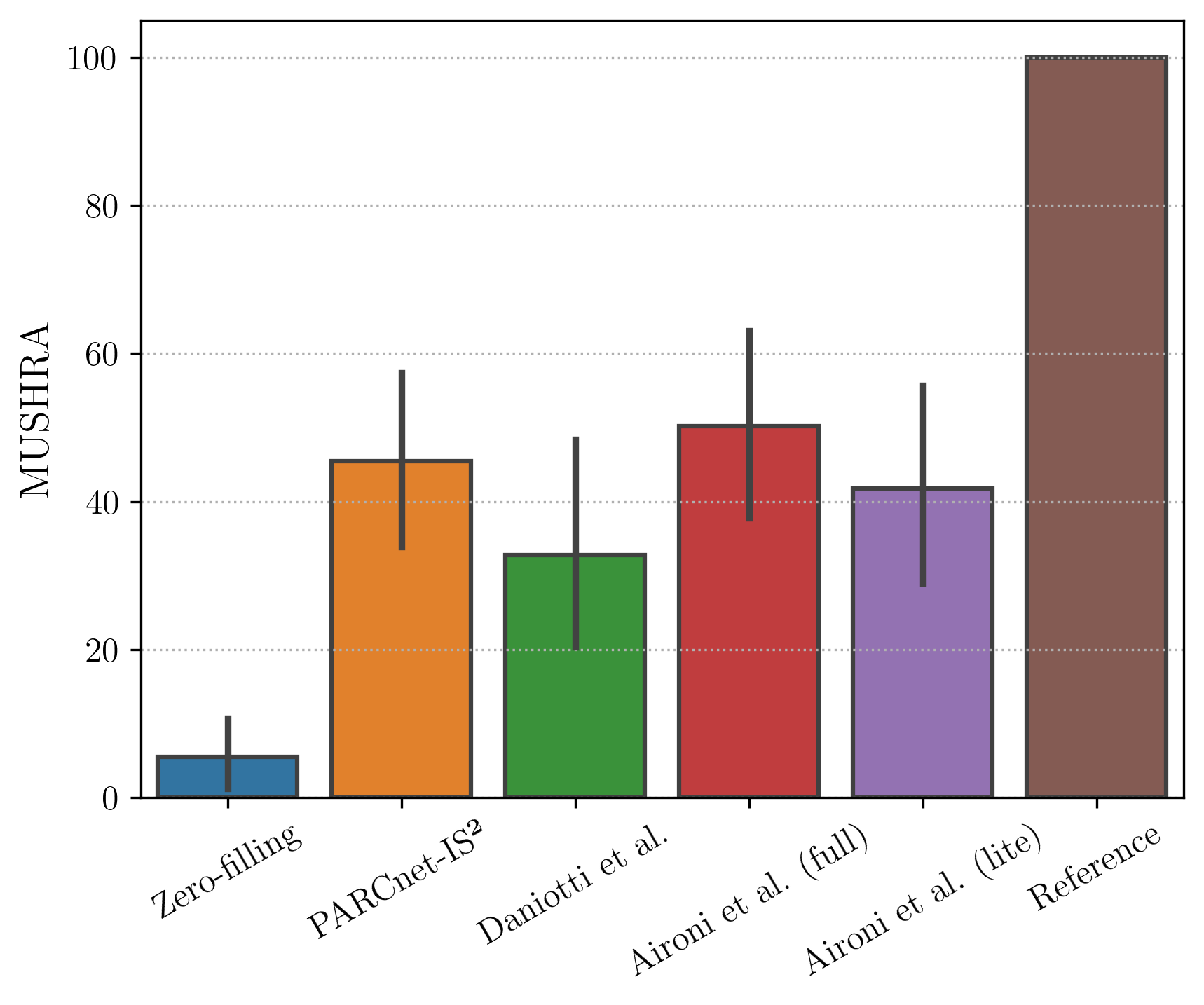}\label{fig:bar-violin2}}\\
    \caption{Average scores and $95\%$ confidence intervals of the individual trials in the MUSHRA test (cont.)}
\end{figure*}

%\null
%\vfill

\end{document}